\def\lncs{0}

\ifnum\lncs=1
\documentclass[9.5pt, conference, compsocconf]{IEEEtran}
\else
\documentclass[10pt]{article}
\fi

\usepackage{amsthm}

\newtheorem*{theorem*}{Theorem}
\newtheorem{definition}{Definition}

\makeatother
\usepackage{cite}
\usepackage{amsmath,amssymb,amsfonts}
\usepackage{amsthm}

\usepackage{pdfsync}
\usepackage{flushend}
\usepackage{mathtools}
\usepackage{algorithmic}
\usepackage{graphicx}
\usepackage{textcomp}
\usepackage{xcolor}
\usepackage{cite}
\usepackage{hyperref}
\usepackage{csvsimple}
\usepackage{subfigure}
\usepackage{adjustbox}
\usepackage{multirow}

\usepackage{tikz}
\usetikzlibrary{shapes,fit,calc,arrows}
\usetikzlibrary{shapes.geometric,shapes.symbols}
\usetikzlibrary{decorations.pathmorphing}
\usepackage{rotating}                   

\def\BibTeX{{\rm B\kern-.05em{\sc i\kern-.025em b}\kern-.08em
    T\kern-.1667em\lower.7ex\hbox{E}\kern-.125emX}}
\usepackage[textsize=tiny]{todonotes}          
\ifnum\lncs=0
\usepackage[left=1in,right=1in,top=1in,bottom=1in]{geometry}
\fi

\newcommand{\bayes}{\ensuremath{\mathsf{Bayes}}}
\newcommand{\exploit}{\ensuremath{\mathsf{ex}}}
\newcommand{\Exploit}{\ensuremath{\mathsf{EX}}}
\newcommand{\ExploitN}{\ensuremath{\mathsf{EX_n}}}
\newcommand{\ExploitV}{\ensuremath{\mathsf{EX_v}}}
\newcommand{\ExploitAND}{\ensuremath{\mathsf{EX_{AND}}}}
\newcommand{\ExploitOR}{\ensuremath{\mathsf{EX_{OR}}}}
\newcommand{\Capab}{\ensuremath{\mathsf{C}}}
\newcommand{\Impact}{\ensuremath{\mathsf{Impact}}}
\newcommand{\Risk}{\ensuremath{\mathsf{Risk}}}
\newcommand{\Reach}{\ensuremath{\mathsf{Reach}}}
\newcommand{\Path}{\ensuremath{\mathsf{Path}}}
\newcommand{\Hybrid}{\ensuremath{\mathsf{Hybrid}}}
\newcommand{\Pred}{\ensuremath{\mathsf{Pred}}}

\newcommand{\Succ}{\ensuremath{\mathsf{Succ}}}
\newcommand{\start}{\ensuremath{\mathsf{Start}}}
\newcommand{\src}{\ensuremath{\sigma}}
\newcommand{\sink}{\ensuremath{\mu}}
\newcommand{\rev}[1]{{\color{blue} #1}} 

\newcommand{\fashion}{\textsc{Fashion}}

\begin{document}

\title{FASHION: Functional and Attack graph Secured\\ HybrId
  Optimization of virtualized Networks}


\ifnum\lncs=0
\author{Devon Callahan\footnote{Email: {\tt devon.m.callahan.mil@mail.mil}. United States Military Academy.} \and Timothy Curry\footnote{Email: {\tt timothy.curry@uconn.edu}. University of Connecticut.} \and Hazel Davidson\footnote{Email: {\tt hazel.davidson@uconn.edu}. Microsoft.} \and  Heytem Zitoun\footnote{Email: {\tt h.zitoun@gmail.com}. University Cote d'Azur - CNRS.} \and Benjamin Fuller\footnote{Email: {\tt benjamin.fuller@uconn.edu}. University of Connecticut.} \and Laurent Michel\footnote{Email: {\tt laurent.michel@uconn.edu}. University of Connecticut.}}
\else
\author{
\IEEEauthorblockN{Devon Callahan\IEEEauthorrefmark{1}, Timothy Curry\IEEEauthorrefmark{2}, Hazel Davidson\IEEEauthorrefmark{3}, Heytem Zitoun\IEEEauthorrefmark{4}, Benjamin Fuller\IEEEauthorrefmark{2}, Laurent Michel\IEEEauthorrefmark{2}}
    \IEEEauthorblockA{\IEEEauthorrefmark{1}United States Military Academy,
    devon.m.callahan.mil@mail.mil.  ORC ID: }
    \IEEEauthorblockA{\IEEEauthorrefmark{2}University of Connecticut, 
    \{timothy.curry, benjamin.fuller,laurent.michel\}@uconn.edu.  }
    \IEEEauthorblockA{\IEEEauthorrefmark{3}Microsoft, 
    hazel.davidson@uconn.edu.}
    \IEEEauthorblockA{\IEEEauthorrefmark{3}Universit\'{e}  C\^{o}te d'Azur - CNRS, h.zitoun@gmail.com\\ORC IDs: 0000-0001-7240-4942, 0000-0002-5772-7468,  0000-0002-9521-8330, \\0000-0002-5772-7468, 0000-0001-6450-0088, 0000-0001-7230-7130}
    }
\fi


%

\maketitle

\begin{abstract}
Maintaining a resilient computer network is a delicate task with conflicting priorities. Flows should be served while controlling risk due to attackers. 
Upon publication of a vulnerability, administrators scramble to manually mitigate risk while waiting for a patch.

We introduce $\fashion$: a linear optimizer that balances routing flows with the security risk posed by these flows.  
$\fashion$ formalizes routing as a multi-commodity flow problem with side-constraints.  $\fashion$ formulates security using  two approximations of risk in a probabilistic attack graph 
(Frigault \emph{et al.}, Network Security Metrics 2017).
$\fashion$'s output is a set of software-defined networking rules consumable by Frenetic (Foster \emph{et al.}, ICFP 2011). 

We introduce a topology generation tool that creates data center network instances including flows and vulnerabilities.  $\fashion$ is executed on instances of up to $600$ devices, thousands of flows, and million edge attack graphs. Solve time averages $30$ minutes on the largest instances (seconds on the smallest instances).  To ensure the security objective is accurate, the output solution is assessed using risk as defined by Frigault \emph{et al.}

$\fashion$ allows enterprises to reconfigure their network in response to changes in functionality or security requirements.

\end{abstract}


\section{Introduction}
\ifnum\lncs=0
Network engineers rely on network appliances to assess the network
state (load, good and bad data flows, congestion, etc.)  and
public vulnerability databases and security appliances to understand  risk. 
Network engineers have to \emph{integrate} both sources to \emph{assess} the overall risk
posture of the network and \emph{decide} what to do.

Due to the intractability of this job, vulnerabilities exist in
enterprise networks for long periods: recent work found it can take
over 6 months to achieve 90\% patching~\cite{kotzias2019mind} (similar
findings in prior studies~\cite{schryen2011open}).  Furthermore, some
vulnerabilities are publicly disclosed before patches are available or
tested, creating a vulnerability window where patching cannot help.
The goal of this work is adjust the network in this vulnerability
window to mitigate risk and maximize functionality.
\fi

(Probabilistic) Attack graphs~\cite{Poolsappasit2011} are labeled
transition systems that model an adversary's capabilities within a
network and how those can be elevated by transitioning to new states
via the exploitation of vulnerabilities (e.g., a weak password, a bug
in a software package, the ability to guess a stack address, etc.). 
We focus on risk that is due to network configuration. See,
for example, the TREsPASS project~\cite{lenin2016technology} for how to incorporate other aspects
of risk such as user error.

There are two parts to running an attack graph analysis, the first is scanning
the network to identify vulnerabilities and mapping them to known hosts in the
network. The first stage can be performed by combining vulnerability scanners
such as Nessus and vulnerability databases such as CVE~\cite{website:CVE}. The
second is taking these vulnerabilities and running some network wide analysis to
understand risk (such as \cite{Dewri2007,Khouzani2019a}). For example, attack
graphs can discover paths that an adversary may use to escalate his privileges
to compromise a given target (e.g., customer database or an administrator
account). We focus on this second stage and how to act on such analysis.

While modern attack graph engines
can issue recommendations~\cite{Dewri2007,sawilla2008identifying,ingols2009modeling,huang2011distilling,Poolsappasit2011,aslanyan2015pareto,pieters2014trespass}, they do
not account for the loss in functionality (i.e., the collateral
damage) that they induce. Furthermore, recommendations must be
implemented manually which increases response time.
A more desirable playbook for coping with emerging threats is:
\begin{enumerate}
\itemsep0em
\item Update security information (in response to a new vulnerability~\cite{website:CVE}), 
\item Generate an attack graph,
\item Derive recommendations from the resulting graph, and 
\item Implement and deploy a set of network rules.
\end{enumerate}

\noindent
The challenges to deliver this vision are threefold: \begin{enumerate}
\itemsep0em
\item evaluate
attack graphs quickly; \item properly route benign flows and drop
risky flows and; \item quickly and transparently deploy changes.
\end{enumerate}

\paragraph{Our contribution} 
Our contribution is an optimization framework called $\fashion$
(Functional and Attack graph Secured HybrId Optimization of
virtualized Networks).  $\fashion$ considers both functionality and
security when deciding how to configure the network. The
\emph{functional} layer is responsible for routing network flows.  It treats network traffic as a multi-commodity
data flow problem. It ensures
that any routing solution carries each flow from its source to its
destination, respects link capacities and network device throughputs,
and satisfies the required demand.  Completeness for the functional layer is the driving factor behind our use of optimization (as opposed to 
machine learning techniques).

Prior work used optimization for attack graph
recommendations~\cite{Dewri2007, Kordy2014}, but there is no analysis scalable to mid size networks (100s of devices) conducive to repeated evaluation on related graphs. $\fashion$ must quickly consider many functionality and
security considerations, meaning that state of the art recommendation
engines are too slow.  

We
introduce security metrics (Section~\ref{sec:attack graphs}) which can be evaluated using binary integer
programming to deliver quick calculation of risk on related
networks. The \emph{security} layer integrates the risk of a
configuration to create a joint model between the two layers.  

This
joint model proposes a network configuration that maximizes functionality while minimizing risk (as described by the attack graph).  
The $\fashion$ model is open source and available at~\cite{fashion-repo}.

\paragraph{System Architecture} 
We focus on software defined networks (SDN). \fashion{}'s output is,
for each flow, the route to serve it, or a decision for how and where
to block it. 
This output can be ingested through the APIs of SDN
controllers. 
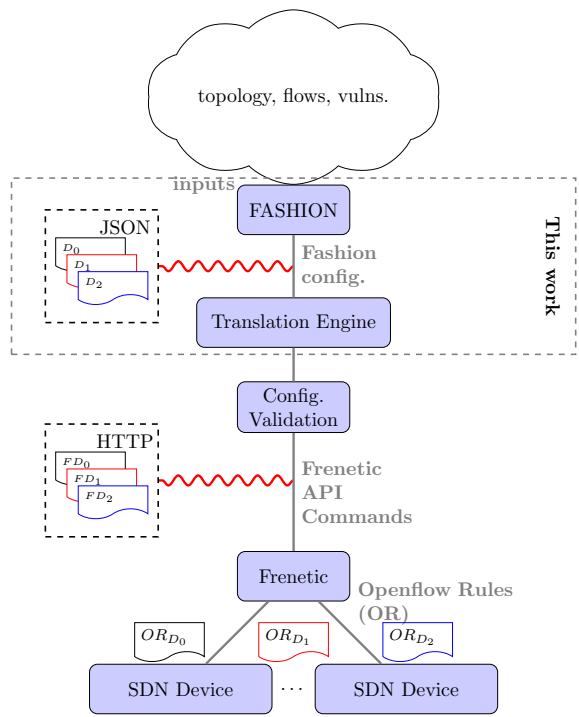
\begin{figure}[t]
\centering

\tikzstyle{block} = [rectangle, draw, fill=blue!20,
    text width=5em, text centered, rounded corners, minimum height=2.5em]
\tikzstyle{line} = [draw, very thick, color=black!50]
\tikzstyle{doc} = [tape,tape bend top=none,text width=1cm,draw,font=\tiny]

\scalebox{.75}{
\begin{tikzpicture}[scale=2, node distance = 2cm, auto,decoration={snake}]
     \node [cloud, draw,cloud puffs=10,cloud puff arc=90, aspect=2, inner ysep=1em] (env) {topology, flows, vulns.};
    \node [block, below=15pt of env] (fashion) {FASHION};
    \node [doc] at (-1.8,-1.50) (d0) {$D_0$};
    \node [doc,fill=white,draw=red] at (-1.7,-1.65) (d1) {$D_1$};
    \node [doc,fill=white,draw=blue] at (-1.6,-1.80) (d2) {$D_2$};
    \node [draw=black,dashed,thick,minimum width = 2cm,minimum height=2cm,fit=(d0) (d1) (d2)] (config) {};
    \path [line,decorate,color=red] (0,-1.65) -- (config);
    \node at (config.north) [below right, inner ysep=2mm, inner xsep=-.5mm]  {JSON};
    %
    \node [block, below of=fashion,text width=3.2cm] (translation) {Translation Engine};
        \node [block, below of=translation,node distance = 1.5cm] (validation) {Config.\\ Validation};
    \node [block, below of=validation,node distance = 3cm] (frenetic) {Frenetic};

    \node [doc] at (-1.8,-3.25) (fd0) {$FD_0$};
    \node [doc,fill=white,draw=red] at (-1.7,-3.4) (fd1) {$FD_1$};
    \node [doc,fill=white,draw=blue] at (-1.6,-3.55) (fd2) {$FD_2$};
    \node [draw=black,dashed,thick,minimum width = 2cm,minimum height=2cm,fit=(fd0) (fd1) (fd2)] (fconfig) {};
    \path [line,decorate,color=red] (0,-3.4) -- (fconfig);
    \node at (fconfig.north) [below right, inner ysep=2mm, inner xsep=-1mm]  {HTTP};
    \node [draw=gray,dashed,thick,minimum width = 10cm,fit=(fashion) (translation)] (us) {};
    \node [below of=frenetic,node distance = 2cm] (sdn0) {$\dots$};
    \node [block, left of=sdn0,node distance = 2cm,text width=3cm] (sdn1) {SDN Device};
    \node [block, right of=sdn0,node distance = 2cm,text width=3cm] (sdn2) {SDN Device};
    \path [line,dashed] (env) -- node [text width=2cm] {\textbf{inputs}} (fashion);
    \path [line] (fashion) -- node [text width=2cm] {\textbf{Fashion\\config.}}  (translation);
    \path [line] (validation) -- node [text width=2cm] {\textbf{Frenetic\\ API\\ Commands}} (frenetic);
    \path [line] (frenetic) -- (sdn1);
    \path [line] (translation) -- (validation);
    \path [line] (frenetic) -- node [text width=3.2cm, above right] {\textbf{Openflow Rules}}(sdn2);
    
    \node at (2.3,-1.5) [rotate=-90] {\textbf{This work}};
    
    \node [doc] at (-1.1,-5.03) (sd0) {\footnotesize $OR_{D_0}$};
    \node [doc,draw=red] at (0,-5.03) (sd1) {\footnotesize $OR_{D_1}$};
    \node [doc,draw=blue] at (1.1,-5.03) (sd2) {\footnotesize $OR_{D_2}$};
\end{tikzpicture}
}
  \caption{$\fashion$ takes as input the existing physical topology, required flows (including bandwidth and value), and vulnerabilities.  The $\fashion$ optimizer outputs a JSON file describing how to route flows in the network (and flows that should be blocked).  This JSON is fed to our Python Frenetic Translation Engine which interacts with the Frenetic controller to create the necessary Openflow tables in the SDN devices.}
  \label{fig:pipeline}
  \vspace{-.15in}
\end{figure}
Specifically, $\fashion$'s output interfaces with the Frenetic
controller~\cite{foster2011frenetic}. Figure~\ref{fig:pipeline} shows
how $\fashion$ fits into a network deployment pipeline.  

We assume
\emph{default deny} routing where only desirable flows 
are carried to their destination.  This corresponds to all extraneous flows in
the network not being served.  This places $\fashion$ in the region where there is a sharp 
tradeoff between functionality and security.  
That is, any reduction in risk demands a reduction in functionality.  
We also assume devices use \emph{source specific routing}~\cite{boutier2015source} which allows packets to be routed
based on the source, destination pair rather than just the destination.  This allows
more flexible decisions, making it easier to respect capacity.
While we consider routing between individual hosts, 
$\fashion$ can also optimize functionality and security
between subnets. Similarly, equivalence classes, used in previous
attack graph research, can also increase scaling~\cite{ingols2009modeling}.

As shown in Figure~\ref{fig:pipeline}, one can use existing tools to validate configurations resulting from \fashion{} before deploying.  If these configuration validations do not pass it may be possible to incorporate their feedback.  If the tool outputs an important flow that is not being
served, the value of this flow can be increased in the configuration.
If the tool outputs a flow that \emph{should not} be served, the risk
of the endpoint can be increased in the configuration.  

\paragraph{Results}
We evaluate three aspects of $\fashion$ 
\begin{enumerate}
\item the feasibility of solving the resulting model
\item if \fashion{} produces good configurations, as measured by prior metrics
  that are too slow for inclusion in an optimization model but can
  \emph{validate the configuration}, and 
\item if $\fashion$ can work \emph{online}, responding to changes in the
    network requirements, we have two questions here:
    \begin{enumerate}
    \item Is solving on related instances faster than starting from scratch, and
    \item Can \fashion{} produce solutions that minimize the change of the
      resulting configuration to minimize the cost of reconfiguration?
\end{enumerate}

\end{enumerate}
The third question of whether \fashion{} works online is crucial to the
  vision of quickly responding to network changes (whether they are new flows or
  changes in vulnerabilities). To study these questions, we use data-center
networks topologies which frequently use virtualized
networking~\cite{Al-Fares:2008:SCD:1402958.1402967}. We use synthetic data for
network flows and vulnerabilities whose statistical properties are drawn from
prior studies. We are unaware of any available dataset for attack graphs. Our
generation tools are open-source~\cite{fashion-repo}. Before deployment in any
real system, the model solve time and solution quality should be measured with
respect to real network demands. These are important pieces of future work.

Timing experiments are discussed in Section~\ref{sec:eval}.
$\fashion$ usually outputs a configuration in under $30$ minutes for the largest tested instances.

Like many integer programming models, the \fashion's objective is a weighted sum, in this case, of a functionality and a security objective. Our functionality objective is the weighted fraction of delivered flows. The main goal of evaluation is to understand whether the security objective effectively models risk.  To understand this question, we report on solution quality using an existing attack graph analysis by Frigault et al.~\cite{frigault2017measuring} that is too slow to be used in the optimization engine.  Recall that by assuming default deny routing, increasing security necessarily means decreasing functionality.  Our main finding (see Section~\ref{ssec:results}) is that this tradeoff is observed for the underlying metric of Frigault et al.~\cite{frigault2017measuring}.  That is, in all observed experiments, increasing weight on the security objective decreases the risk as measured by Frigault et al.'s metric. (This monotonicity is of course present in the optimization objective.)
\subsection{Toy Example}
\label{sec:toy example}
A major component of this work is efficiently modeling risk as measured by an attack graph.
Towards this end, we describe a toy attack graph example. Actual attack graphs are built by combining network scans and vulnerability databases~\cite{chu2010visualizing}. 

\ifnum\lncs=0
The objective of the
framework is to produce decisions to configure the
network devices (switches, firewalls, etc.) while balancing functional and
security needs. 
In this work we consider routing decisions on flows only (including blocking a flow).
\fi

\begin{figure}[t]
\centering
\includegraphics[scale=.65]{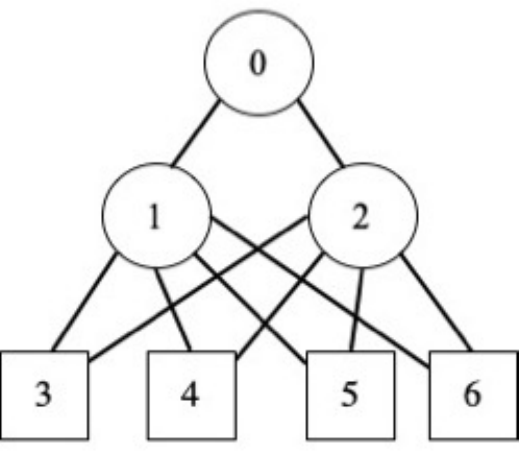}
\caption{
The physical topology of the network and the set of available
network links that can be used by the optimization framework.
}
\label{fig:configurations}
\end{figure}

Figure~\ref{fig:configurations} illustrates the physical layer of the
network. The following conventions are used: circle are
  switches, square are hosts, and black lines are physical
  connections.
    It features 3 SDN appliances, nodes $0$, $1$ and $2$
that route traffic as well as block it (act as firewalls). The toy
network features 4 hosts, $3$ through $6$.

\begin{table}[t]
%
\centering
\begin{tabular}{lll}
  \hline
  src $\rightarrow$ dst & type & value \\ 
  \hline
0 $\rightarrow$ 3 &  A &  \$\$\$ \\ 
3 $\rightarrow$ 4 &  A &  \$ \\ 
3 $\rightarrow$ 4 &  B &  \$ \\ 
3 $\rightarrow$ 5 &  A &  \$\$ \\ 
3 $\rightarrow$ 5 &  B &  \$ \\ 
5 $\rightarrow$ 6 &  A &  \$\$ \\ 
   \hline
\end{tabular}
\quad
\begin{tabular}{l |  l r}
  \hline
Preconditions & post \\
  \hline
($3$,$A$)  & ($3$,$\mathtt{Code}$)  \\ 
($5$,$A$)  & ($5$,$\mathtt{Code}$)  \\ 
($6$,$A$), ($5$,$\mathtt{Code}$)  & ($6$,$\mathtt{Code}$)  \\ 
($4$,$B$), ($3$,$\mathtt{Code}$)  & ($4$,$\mathtt{Code}$)  \\ 
   \hline 
\end{tabular}

\vspace{1mm}
\caption{Flows (left) and exploits (right) in the network in Figure~\ref{fig:configurations}.}
\label{tab:exploits}
\label{tab:routing}
\vspace{-.3in}
\end{table}

Table~\ref{tab:exploits} shows the required flows and security vulnerabilities in the network.
This physical network must be configured to serve traffic
demands. Table~\ref{tab:routing} shows a collection of data flows in
the form $s\rightarrow t$ which a flow from node $s$ to node $t$. Each data flow has a type (here $A$ or $B$). Each
data flow carries an economic value shown by the number of $\$$
signs. 

Table~\ref{tab:exploits} also shows exploits. 
A
pair $(h,p)$ indicates the adversary has privilege $p$ on host
$h$ where privileges are in $\{A, B, \mathtt{Code}\}$.  The attacker can have three types of privileges, the ability to: 1) send traffic of type $A$, 2) send traffic of type $B$ or 3) execute code, denoted as $\mathtt{Code}$. Exploits have one or more pre-conditions
that must be met to achieve the new privilege. Privilege level $\mathtt{Code}$ indicates the ability to execute code on that host.  For the toy example, we assume all exploits have probability $1$
of being achieved if preconditions are met. 

\begin{figure}[t]
\centering

\tikzstyle{capab} = [ellipse, draw, text width=5em, text centered,minimum height=2em]
\tikzstyle{exploit} = [diamond,aspect=2,draw,text width=1cm,minimum height=1em,text centered,draw=red]
\tikzstyle{net} = [rectangle,draw=green,text width=3em,text centered,minimum height=2em]
\tikzstyle{line} = [draw,->,>=stealth',line width=0.5mm, color=black]
\tikzstyle{lineB} = [draw,->,line width=0.5mm, color=blue!50]
\tikzstyle{lineE} = [draw,->,line width=0.5mm, color=red]

\scalebox{.54}{
\begin{tikzpicture}[node distance = 2cm, auto]
  \node[capab] at (0,0) (start) {(0,A)};
  \node[net] at (0,-2) (net) {NET};
  \node[capab] at (0,-4) (cap30) {(3,A)};
  \node[net] at (-6,-6.5) (net10) {NET};
  \node[net] at (-3,-6.5) (net0) {NET};
  \node[exploit] at (0,-6.5) (exp0) {exploit 0};
  \node[net] at (3,-6.5) (net1) {NET};
    \node[net] at (6,-6.5) (net11) {NET};
\node[capab,below of=net10] (cap4A) {(4,A)};
\node[capab,below of=net0] (cap4B) {(4,B)};
\node[capab,below of=exp0] (cap31) {(3,$\mathtt{code}$)};
\node[capab,below of=net1] (cap5A) {(5,A)};
\node[capab,below of=net11] (cap5B) {(5,B)};
  \coordinate (CENTER0) at ($(cap4B)!0.5!(cap31)$);
  \node[exploit,below of=CENTER0] (exp3) {exploit 3};
  \node[net,below left of =cap5A] (net2) {NET};
  \node[exploit,below right of=cap5A] (exp1) {exploit 1};
\node[capab,below of=exp3]  (cap41) {(4,$\mathtt{Code}$)};
\node[capab,below of=net2]  (cap60) {(6,A)};
\node[capab,below of=exp1]  (cap51) {(5,$\mathtt{Code}$)};
  \coordinate (CENTER) at ($(cap60)!0.5!(cap51)$);
  \node[exploit,below of = CENTER] (exp2) {exploit 2};
  \node[capab,below of=exp2] (cap61) {(6,$\mathtt{Code}$)};
  
\path[line]
(net2) -- (cap60)
(cap5A) -- (net2)
(start) edge node  {} (net)
(net) edge node  {a} (cap30)
(net0) edge [bend left=-10] node {b} (cap4B)
(cap30) edge [bend left=-10] node  {} (net0)
(cap30) edge [bend left=-10] node  {} (net10)
(net10) edge [bend left=-10] node  {} (cap4A)
;
\path[line]
(net1) edge [bend left=10] node {c} (cap5A)
(net11) edge [bend left=10] node {} (cap5B)
(cap30)  edge [bend left=10] node  {}  (net11)
(cap30)  edge [bend left=10] node  {}  (net1);

\path[lineE] (cap30) -- (exp0);
\path[lineE] (exp0) -- (cap31);
\path[lineE] (cap31) -- (exp3);
\path[lineE] (cap4B) -- (exp3);
\path[lineE] (exp3) -- (cap41);
\path[lineE] (exp2) -- (cap61);
\path[lineE] (cap60) -- (exp2);
\path[lineE] (cap51) -- (exp2);
\path[lineE] (exp1) -- (cap51);
\path[lineE] (cap5A) -- (exp1);

\end{tikzpicture}
}
\caption{Attack graph for network in Figure~\ref{fig:configurations}.}
\vspace{-.15in}
\label{fig:functional:ag}
\end{figure}
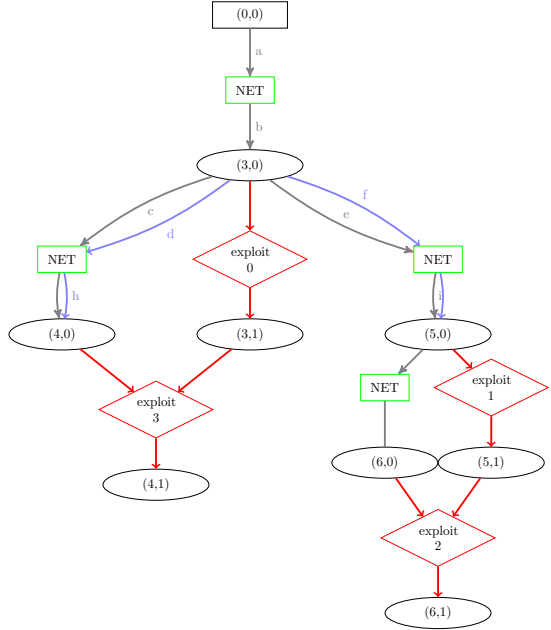

 Figure~\ref{fig:functional:ag} conveys the attack graph for this
 network which is built by combining the flows and exploits in Table~\ref{tab:exploits}.\footnote{This representation is known as an attack dependency graph, see Section~\ref{sec:attack graphs}.}  Figure~\ref{fig:functional:ag} shows how an attacker can gain privileges using either flows that exist in the network or exploits.  Figure~\ref{fig:functional:ag} uses the following
     conventions: 1) circle nodes are (host,privilege)
   states, 2) green square NET nodes represent network reachability, 3)
   diamond nodes are exploits, 4) black arcs correspond to network
   connection, 5) incoming red links are
   precondition states of exploits, and 6) outgoing red links are
   postcondition states of exploits. 
      
To illustrate, there is a required flow between host $0$ and host $3$ of type $A$. So precondition $(0, A)$ yields postcondition $(3, A)$.  This is shown as a NET connection. A vulnerability on host $3$ transforms the $(3, A)$ precondition into the $(3, \mathtt{Code})$ postcondition.   This is shown in the exploit 0 diamond in Figure~\ref{fig:functional:ag}. Note that exploits $2$ and $3$ require the attacker to have two preconditions to gain the new postcondition.


The goal of \fashion{} is to configure SDN rules to minimize the impact of known attacks without unnecessarily sacrificing functionality.  Specifically, each output of \fashion{} is a set of routing tables for each SDN appliance (including explicit firewalls). Here we focus on \fashion's decision of which flows to serve, this corresponds to deciding which if any edges of the attack graph to cut. 

We now discuss how \fashion{} would operate in the setting that functionality or security was considered followed by a joint setting where both were considered.
\begin{itemize}
\item \textbf{Functionality Only} All flows would be routed in the network and the attacker would be able to gain $\mathtt{Code}$ privilege on nodes $3$ through $6$.
\item \textbf{Security Only} The most effective way to prevent the attacker from gaining $\mathtt{Code}$ privilege is to have node $0$ add a firewall rule blocking all traffic destined for node $3$
of type $A$.  This corresponds to cutting edge $a$ in Figure~\ref{fig:functional:ag}.
In this configuration, the internal traffic proceeds
unabated. However, the high  value of flow
$(0,A) \rightarrow (3,A)$ was not respected doing great damage to the
value of the network service.
\item \textbf{Balanced Configuration} A balanced configuration continues to serve flow $(0, A) \rightarrow (3,A)$ because of its economic value.  However, one can prevent the adversary from achieving $\mathtt{Code}$ on machines $4, 5$ and $6$ by sacrificing two of the flows in Table~\ref{tab:exploits}.  Specifically, the SDN devices add block rules to prevent routing the flows $(3, A)\rightarrow (4,B)$ and $(3, A) \rightarrow (5,A)$.  This corresponds to cutting edges $b$ and $c$ in the attack graph. These two flows are chosen because of their low value in comparison to the amount of $\mathtt{Code}$ privileges they prevent the attacker from achieving. 
\end{itemize}

In the above, we treat \fashion{} as just deciding which flows to serve, the actual framework also creates and places SDN rules on all SDN devices. That is, 1) for every required flow in the network \fashion{} creates routes through the topology in Figure~\ref{fig:configurations}  and 2) for blocked flows, \fashion{} ensures that traffic is routed to a firewall that drops the traffic.


\noindent
\paragraph{Organization}
The organization of this work proceeds as follows.  Section~\ref{sec:rel work} describes related work on the major system components.  
 Section~\ref{sec:attack graphs} introduces background on
attack graphs and on the measures we will optimize over, Section~\ref{sec:model} 
documents the optimization model, Section~\ref{sec:eval} evaluates $\fashion$, and 
Section~\ref{sec:conclusion} concludes. 

\section{Related work}
\label{sec:rel work}

Attack graphs are not a panacea, they require effort to find the necessary inputs (e.g. using a vulnerability scanner)~\cite{Aksu2018,Piwowarski2005} and it is unclear how to implement the resulting recommendations \cite{Lippmann2005} which usually correspond to some arc to
cut~\cite{Bistarelli2006}.
However, cutting an arc without major
damage is hard for a human analyst.  $\fashion$ is designed to address
this problem.  SDN
provides the perfect opportunity for integration with attack graphs.
SDN offers a centralized control and holistic view of the network no
longer requiring external scanning tools to discover reachability
data~\cite{fayaz2015bohatei}. 

Attack graphs grow much faster than the
underlying network (see Table~\ref{tab:benchmarks}), which affects analysis time.  As we discuss in Section~\ref{sec:attack graphs}, we use an attack graph representation that is amenable to larger networks.  Assessment of risk in an attack graph is a complex task.  



Researchers have proposed high-level SDN programming languages in
order to efficiently express packet-forwarding policies and ensure
correctness when dealing with overlapping
rules~\cite{foster2011frenetic,reich2013modular}.  These languages
focus on parallel and sequential composition of policies to ensure
modularity while providing correctness guarantees. When
\fashion{} proposes a set of new rules for the controller it is important
to change routing while minimizing loss~\cite{reitblatt2012abstractions}.

To aid in network configuration, research tools
assess network reachability~\cite{khurshid2012veriflow}, network security
risk~\cite{schneier1999attack,yu2018deploying}, and link contention \cite{wang2011openflow,skowyra2014verification}.  These tools assess the quality of a configuration with respect to 
a single property and do not provide recommendations.

Recent tools generate network configurations from a set of functional and security 
requirements~\cite{Subramanian2017,El-Hassany,Majumdar2017,Narain2008,Beckett2016}. Captured security requirements include
IPSec tunneling, allowing only negotiated packet flows, and ensuring identical rule sets on all firewalls.

Our work can be seen as unifying three recent works, one by Curry \emph{et
al.}~\cite{curry2019docsdn}, one by Frigault et al.~\cite{Frigault2008b,frigault2017measuring} and another by Khouzani \emph{et
al.}~\cite{Khouzani2019a}.  
Curry \emph{et al.} proposed an optimization framework for deciding on a
network configuration based on the given desired network
functionality of data flows and the underlying physical network.
Curry \emph{et al.}  produce network configurations
that meet all demands while blocking adversarial traffic. Each network node has an input risk and nodes assume a fraction
of the risk of any node with which they share a path.  Their
risk measure does not consider adversaries that pivot in the network.  To address this shortcoming we consider two risk measures in prior work, an declarative measure of Frigault et al. and an optimization based measure of Khouzani et al.

Frigault et al.~\cite{Frigault2008b,frigault2017measuring} present a risk
  assessment metric that assumes independence of compromise events.
  Specifically, if an adversary has two possible ways to reach some node, their
  total probability of reaching that node is the sum of the probability of the
  two paths minus the product of the two probabilities. If the adversary must
  obtain two capabilities to capture a node, the probability is the product of
  the two probabilities. This calculation is precise if compromise probabilities
  are independent. Frigault et al. present no timing information on their
  algorithm only scaling to six node attack graphs. Looking forward, we use this
  algorithm as a baseline for assessing the risk in our candidate
  configurations. Even with the assumption of independence, Frigault et al.'s
  metric can take minutes to compute on networks with 200 devices (see
  Section~\ref{sec:conclusion}), this timing is based on an our implementation
  of this algorithm which has been open-sourced with the rest of \fashion{}.

Khouzani \emph{et al.}~\cite{Khouzani2019a} created an optimization engine designed to
minimize security risk as represented by an attack graph.  They show
how to formulate the most effective path of an attack graph using a
linear program.  Their functionality view is limited to imparting an explicit numeric
functional \emph{cost} to each remedial action.  It is
unclear how to create these costs.
Additionally, the size of Khouzani et al.'s optimization model grows exponentially in the size of the network, making it intractable for all put the smallest networks.  

 The primary technical gap in integrating these three works is
  \emph{creating an optimization model that effectively assesses risk in an
    attack graph representation that scales to moderate size networks.}

Chia's~\cite{chia2020spar} subsequent model and evaluation are nearly identical to ours, so we do not compare with their results.




\section{Attack Graphs}
\label{sec:attack graphs}
$\fashion$'s goal is to balance the functionality and security needs of the network.  Functionality needs are relatively straightforward to state: a set of desired network flows that should be carried in the network while respecting link constraints such as bandwidth.  

For security needs, we use the abstraction of \emph{attack graphs}.  
Attack graphs model paths that an attacker could use to penetrate a network~\cite{schneier1999attack,Sheyner2002,Ingols2006,ammann2002scalable}. 
Paths combine network capabilities such as routing and exploitation of a software/hardware vulnerabilities.  
\ifnum\lncs=1
An attack graph assumes an attacker starts at some entry point such as a publicly facing Web page.
\else
An attack graph assumes an attacker starts at some entry point such as a publicly facing Web page 
and through a series of privilege escalations and network device accesses pivots to eventually reach his or her desired destination.  (The technology supports an arbitrary starting point if one wishes to consider insider attacks.)  
\fi

\label{ssec:expanded view}
Two
common graph representations are an attack \emph{dependency} graph and
an attack \emph{state} graph.  In the \emph{dependency} view each node
represents an exploit or a capability in the network and a Boolean formula over prerequisites determines whether an attacker may achieve a capability (see Fig~\ref{fig:functional:ag}).   The main
drawback of this representation is that the analysis of overall
risk is difficult.  In the \emph{state} graph view, each node represents an
attacker's current capabilities.  This representation eliminates cycles which simplifies
analysis.

However, the representation is exponentially larger than the corresponding dependency graph.
Consider a system in which $k$ 
abstract capabilities must be encoded. In a dependency graph, one needs $k$
nodes, one per capability. In a state graph, one needs $2^k$ nodes. Each node
denotes the set of capabilities secured by the attacker and an arc between two
states encodes the acquisition of a capability (one never ``loses'' a 
capability). This exponential blowup is prohibitive even for moderate size
networks~\cite{Homer2009}. Khouzani~\cite{Khouzani2019a} 
requires the state representation to model and minimize risk.
The sheer size of \emph{attack state graphs} 
prompts us to adopt the \emph{attack dependency graphs}
representation.

In the  dependency view,
each node captures an exploit or capability
in the network.  It may be possible to reach a capability using many
different paths.  Furthermore, multiple conditions may be necessary to
achieve this capability, for example, network reachability of a
database machine and a SQL injection attack.   
\ifnum\lncs=0
This is the view presented in Figure~\ref{fig:functional:ag}.  
\fi

Each exploit  has an associated Boolean structure (indicating when the
exploit can be obtained) and a probability (indicating attacker
success rate in carrying out the exploit).  Capability nodes are
annotated with an impact value that signifies the cost of an adversary
achieving that capability (following the NIST cybersecurity framework
guidance~\cite{website:CF}).  
In this work, we focus on exploits that
are AND and OR prerequisites (and not arbitrary Boolean formulae).  Intuitively, an AND node means the attacker must have all prerequisites to achieve the new capability.  An OR node means the attacker must have a single prerequisite to achieve the new capability.

While a dependency representation is compact, risk analysis is delicate.
Even if one assumes that probability associated with each
arc is independent, calculating the overall probability requires
considering all paths, a tall order in the presence of cycles. 
We return to this
problem after introducing prior work
on quickly evaluating attack graphs and formalizing the state graph representation.

\paragraph{Prior work on evaluating related attack graphs}
\label{sec:dynamic graph rel work}
The ability to perform this analysis quickly is critical to utilize
attack graphs in our optimization framework.  The ability to
regenerate the attack graph multiple times at decreased cost has been
addressed recently in related contexts.

Almohri \textit{et al.}~\cite{Almohri2016} considered an attack graph
setting where the defender has incomplete knowledge of the network.
An example source of stochasticity is mobile device movement. Their
attack graph model can capture risk subject to such stochasticity
sources. This is an orthogonal consideration to our work that could be 
folded into $\fashion$'s framework as part of future work.


Frigault et al.~\cite{Frigault2008b,frigault2017measuring} and Poolsappasit et
al.~\cite{Poolsappasit2011} argue that it is inaccurate to measure
probabilities with a fixed probability of exploit.  They argue that
factors such as patch  availability will decrease the threat while
widespread distribution of vulnerability details may increase the
threat.  As such, they conduct attack graph analysis where the graph
is static but probabilities can change over time.  
If one is willing to consider a \emph{complete} attack graph
then all changes in the graph can be represented with a change in
probability.  However, this also introduces an exponential blowup 
as one needs to consider an arc between each subset of nodes. 

Khouzani \emph{et
  al.}~\cite{Khouzani2019a} is the only work that considers defensive
actions that can drastically change the attack graph and a formulation
of the risk calculation graph that is amenable to optimization.
However, their model is inherently tied to the state graph
representation.

\subsection{Formalizing the problem}
\label{ssec:formalizing problem}
Building an attack graph requires network reachability information, device software configurations and known exploit information~\cite{website:CVE, website:NVD,website:CVSS,website:CWSS} to generate the graph. Attack graphs are effective in measuring how an attacker would traverse using known vulnerabilities and system state.  One can also consider the implications of a new vulnerability~\cite{ingols2009modeling}.

This section defines the risk measure used as ground truth when evaluating 
$\fashion$ solutions.  The metric is drawn from Frigault \emph{et al.}~\cite{frigault2017measuring}
augmented with impact for each node.  Frigault \emph{et al.}'s metric assumes an
attack graph where the probability of achieving each exploit is
independent.  Since paths in an attack graph often overlap, the
probability of achieving prerequisites of an exploit may not be
independent. This simplifying assumption is often used since
considering correlated probabilities makes the problem significantly
harder~\cite{Homer2013}.

\ifnum\lncs=0
In isolation a misconfigured device which allows unauthorized access may be benign but when coupled with network access to ex-filtrate data or pivot to additional targets the results can be devastating. 
The goal of constructing and analyzing an attack graph is to understand the security posture in total.  An attack graph should allow one to understand defensive weaknesses and critical vulnerabilities in the network.  Since all enterprises have limited budgets, the goal of this analysis is usually to prioritize changes that have the largest impact. 
\fi

 An attack graph $G(N,E)$ is a directed graph. 
 Let $\Exploit = \ExploitN \cup \ExploitV$ be the set of all
  exploits where $\ExploitN$ represents the network reachability exploits (changes that can be made 
  by the SDN controller) and $\ExploitV$ represents the set of
  vulnerabilities. We rely on the following assumptions: 
\begin {itemize}
\item That $\Exploit = \ExploitAND \cup \ExploitOR$,
  i.e., the set of exploits can be partitioned into conjunctive or
  disjunctive nodes.
\item That $\ExploitN \subseteq \ExploitOR $, i.e.,
  reachability is treated as OR between multiple traffic types
  among hosts.
\item Exploits $\exploit \in \Exploit$ are labelled with a probability
  $p(\exploit)$ of exploitability given that all prerequisites have
  been satisfied.  It can be estimated using vulnerability
  databases~\cite{website:CVE,website:NVD,website:CVSS}. 
  \ifnum\lncs=0
  Note that
  $p(\exploit)$ is a component metric, we seek to capture the
  cumulative risk in the network (see discussion
  in~\cite{frigault2017measuring,singhal2017security}).
  \fi
\end{itemize}

Let $\Capab$ be the set of all capabilities in the
  network.  A capability $\Capab$  carries an impact, denoted as
  $\Impact(\Capab)$ which is nonnegative.
 Let $N = \Exploit \cup \Capab$ denote the node set.
 Let $E = R_r \cup R_i$ denote the arc set with
\begin{itemize}
\item $R_r \subseteq \Capab\times\Exploit$ : the
  exploit's prerequisites.
\item $R_i \subseteq \Exploit\times\Capab$ :  
  capabilities gained from an exploit.
\end{itemize} 
There are no arcs in $\Capab \times \Capab$ or $\Exploit \times \Exploit$. 
 Let $\start \subseteq \Capab$ denote a set of capabilities that
  the adversary is believed to have.
For any node $n$, $\Pred(n)=\{v| (v, n) \in E\}$ denotes all its predecessors 
in $G$, while $\Succ(n)=\{v| (n, v)\in E\}$ denotes all of its successors.

\paragraph{Risk without Cycles} First, consider how to compute risk in the absence of cycles, we follow Frigault \emph{et al.}.'s~\cite{frigault2017measuring} methodology which is centered on Bayesian inference, augmented with an impact for
each capability node.

The primary goal is to compute cumulative scores $P(\exploit)$ and
$P(c)$ for each node in $G$. The cumulative scores $P$ are computed using the attack graph and the component score of each exploit, denoted as $p(\exploit)$.These represent the likelihood that an
attacker reaches the specified node in the graph.\footnote{It is
  possible to assign individual component scores $p(c)$ for nodes
  $c\in \Capab$.  In this work we assume that $p(c)=1,
  \:\forall\:c\in\Capab$. That is, we assume that
  all uncertainty in the attacker's success is represented in exploit
  arcs.}  From these scores one may consider the overall
risk:
\begin{equation}
 \label{Risk: exact}
\Risk(G) = \sum_{c\in C} P(c) * \Impact(c)
\end{equation}
$P(\cdot)$ is computed as follows in the acyclic case:

\paragraph{AND nodes}  For AND
nodes, the cumulative score is the product of $P(c)$ for all predecessors and
$p(\exploit)$.  The intuition is that each prerequisite must be achieved and the probabilities are assumed to be independent.
\paragraph{OR nodes} For OR nodes, the cumulative score is the sum of
all predecessors component score minus the product of each pair of
probability scores (using Bayesian reasoning).  All capability nodes
are treated as OR nodes.  For a set $X$, we define the operator
$\bayes: X \rightarrow [0,1]$ as:
\[
\bayes (X) =\begin{cases} \sum_{X'\subseteq X} \left( -1\right)^{|X'|+1} \prod_{x\in X'} p(x) & X\neq \emptyset\\0 & X =\emptyset\end{cases}.
\]
$\bayes(X)$ measures the probability that at least one event in $X$ occurs (events in $X$ is assumed to be independent).

\begin{definition}[Network Risk] \cite[Definition 2]{frigault2017measuring}
\label{def:network_risk}
\textit{Given an acyclic attack graph }$G$
\textit{and any component score assignment function} $p :   \Exploit \longrightarrow\left[0,1\right],$ 
\textit{the cumulative score function} $P: \Exploit \cup \Capab \longrightarrow \left[0,1\right] $ \textit{is defined as} 
\begin{align*}
P(\exploit) &= \begin{cases} p(\exploit) \cdot \prod_{c \in \Pred(\exploit)} P(c) & \exploit  \in \ExploitAND \\ p(\exploit)\cdot  \bayes(\Pred(\exploit)) & \exploit \in \ExploitOR \end{cases}\\
P(c) &=\begin{cases} 1 & c\in \start\\
 \bayes(\Pred(c))& \text{otherwise}.\end{cases}
\end{align*}
\end{definition}

$P(n)$ can be computed for all $n\in N$ as long as $G$ is
acyclic.  This is because $P(n)$ can be computed once $P(e)$ is known for all
$e\in\Pred(n)$.  For all acyclic $G$  there is a topological ordering
that allows this computation (and all topological orderings result in
the same computation).  

Frigault et al.~\cite{frigault2017measuring} present a complex algorithm to handles cycles; it involves repeated, recursive evaluation of network risk on the graph with arcs removed to break cycles. 
\label{ssec:cycles}
Frigault et al.~\cite{frigault2017measuring} observed that:
\begin{enumerate}
\item We only need to measure the probability for the first visit of a
  node.  Consider a cycle with only one incoming arc. The arc closing the cycle can be ignored.
\item Cycles with multiple entry points are more delicate to handle
  and require the removal of an arc from the graph.  The key
  insight is no path that an attacker traverses will actually
  follow the cycle.  Different paths will include subsets of arcs
  from the cycle.
\end{enumerate}

Frigault et al.~\cite{frigault2017measuring} propose the following methodology for
handling cycles with multiple entry points.  Assume
that all nodes that can be topologically sorted have been and their
cumulative probabilities are assigned.  Let $X$ be a cycle with at
least two entry points.  For each entry
point $x\in X$ one can compute $P(x)$ without considering $\Succ(x)$.
While $x$'s successors are important in calculating the overall risk
they do not impact $P(x)$.  Compute a new attack graph $G'$
which has all $\Succ(x)$ removed and use it to calculate $P(x)$.
Importantly, the graph $G'$ may still have cycles which inhibit
computation of $P(x)$ requiring this process to be repeated for all entry points in the cycle.  Once all entry points have their likelihood, the
rest of the cycle can be safely evaluated.


When we use the term $\Risk(G)$ it refers to this full computation. $\Risk$
is too slow to directly use for the security layer (see Section~\ref{sec:eval}).
A Python implementation of Frigault \emph{et al.}'s
algorithm~\cite{frigault2017measuring} has been open-sourced along with the rest of
\fashion~\cite{fashion-repo}.

\subsection{Approximating Risk}
\label{ssec:approx_risk}
To incorporate cumulative risk into an optimization framework, we turn
to approximations of risk that can be linearized.  The risk
calculation presented in Definition \ref{def:network_risk} and its
augmentation for handling cycles is non-linear and has no closed form.
We consider two approximations to serve as proxies called
\emph{Reachability} and \emph{Max attack}.
We defer the evaluation of the quality of our measures to Section~\ref{sec:eval}.

 \subsubsection{Reachability}
We binarize probability of exploitation resulting in a metric called $\Reach$.
\ifnum\lncs=0 
$$P^*(n) = \begin{cases}1 &P(n) > \theta \\ 0 &P(n)\leq\theta\end{cases}$$
The value $\theta$ is a threshold used to determine whether a capability should either 
be ignored or considered attained. We consider $\theta =0$.  Since $p(n)\in 
\{0,1\}$, we can apply standard Boolean linearization techniques to get a tractable 
representation of dynamic risk.

 \begin{equation}
 \label{risk_width}
\Reach(G^*) \overset{def}= \sum_{c\in C} P^*(c) * \Impact(c)
\end{equation} 

\else
That is, any arc in the attack graph with a nonzero probability is assumed to be compromised, so all probabilities are either $0$ or $1$.  Since the probability is binary we can apply standard Boolean linearization techniques to get a tractable 
representation of dynamic risk.
\fi

\ifnum\lncs=0
Utilizing the binary representation of exploits in Equation~\ref{risk_width} is an approximation, measuring how an attacker can impact a target network.  
Since we consider $\theta=0$ this measures the total impact of nodes reachable by the adversary.  
\fi
This is equivalent to calculating the weighted size of the connected
components in $G$ that contains the attacker's starting posture. The first generation of
attack graphs  measured this quantity~\cite{Sheyner2002,Ritchey2000}. This models the \textit{worst case} approach when calculating attacker compromise of network capabilities. Note that one can set some threshold for mapping a probability to $1$ rather than all nonzero probabilities.

This approach does have a weakness when the goal is to jointly optimize functionality and security.  Consider two nodes $a$ and $b$ where $\Impact(a) = 2*\Impact(b)$.  Further suppose, at least $a$ or $b$ must remain in the connected component to satisfy functionality demands.  The reachability metric will prioritize disconnecting $b$.  However, it may be that the attacker is less likely to reach $b$.
This may be the case even if $p(b)\ge p(a)$ (it is possible that $P(a)>P(b)$ due to the likelihood of reaching their predecessors). 

%

\label{ssec:max-risk-path} \subsubsection{Most Likely Path} The second risk measurement we introduce is the attacker's most likely course of action, called $\Path$. This measure is based on Khouzani \emph{et al.}'s \emph{most effective attack measure}~\cite{Khouzani2019a}.  

\newcommand{\pathSet}[1]{\omega_{\src \rightarrow #1}}
\newcommand{\normPact}{\Lambda}
Let $\src$ be the attacker's starting point in the attack graph.
We define $\pathSet{c}$ to be the set of all paths, where a path is sequence of 
arcs 
$(e_1, \dots, e_k)$ such that $e_i \in E$, from 
$\src$ to $c$ in the attack graph. Let $\normPact_{c} \in [0,1]$ be 
the normalized impact of an attacker obtaining capability $c$. Then 
the most effective attack path is defined as follows.

\label{Risk: width}

\begin{equation}
\label{pathRisk: multiple targets}
	\Path(G) = \max \limits_{c \in \Capab} \normPact_{c} 
	\max\limits_{\pathSet{c}} 
	\prod_{e \in 
	\pathSet{c}} p(\Pred(e))
\end{equation}
Instead of having multiple targets, an auxiliary 
target  $\sink$ is considered, that will be the sole target capability of the 
attacker. To do this, arcs from each $c \in 
\Capab$ to an auxiliary exploit $\exploit_{c}$ are introduced, such that 
$p(\exploit_{c}) = \normPact_{c}$. Then we introduce an arc from 
each $\exploit_{c}$ to $\sink$. Doing this, we can reformulate 
equation~\ref{pathRisk: multiple targets} as:
\begin{equation}
\label{pathRisk}
\Path(G) = \max\limits_{\pathSet{\sink}} \prod_{e \in \pathSet{\sink}} 
p(\Pred(e))
\end{equation}
However network defenses can be deployed in order to reduce the 
probabilities of these 
exploits. Let $x_d \in \{0,1\}$ be a binary decision variable 
denoting whether 
a network defense $d$ has 
been deployed. Let $p_{d}(\exploit)$ be the 
reduced probability of exploit $\exploit$ due to network defense $d$. 
With this, the probability of exploit $\exploit$ with 
respect to network defense decision $x_d$ is given by
\begin{equation}
\label{probRed}
	p_{x_d}(\exploit) = p(\exploit) (1-x_d) + p_{d}(\exploit) x_d.
\end{equation}

We will minimize the risk due to the most effective path
risk over all the possible configurations of network defenses
available.  This approach identifies appropriate locations to deploy
network defenses to protect both high value capabilities with
low exploitability as well as lower value, more exploitable assets.  We
will incorporate these defense decisions when defining our
optimization model in Section~\ref{sec:model}.

$\Path$ does not distinguish between a graph with $2$ paths with the same underlying probability and a single path with that probability.  In addition to this inaccuracy (that was present in Khouzani \emph{et al}'s work) working on the attack dependency graph introduces two sources of error:
\begin{enumerate}
\item $\Path$ only counts the impact from the last node on the path.  This is because it is only measuring the probability of a path and the impact is added as a ``last layer'' in the graph.  So it does not distinguish between two paths (of equal probability) where one path has intermediate nodes with meaningful impact.  In the attack state representation each node has the current capabilities of the attack and thus impact of this node set can be added as a last layer.
\item The path used to determine $\Path$ may not be exploitable by an adversary due to 
exploit nodes with multiple prerequisites.  That is, the path may include a node with multiple prerequisites and the 
measure only computes the probability of exploiting a single prerequisite. In the attack 
state representation there are no nodes with multiple prerequisites so this problem does not arise.
\end{enumerate}

\subsubsection{Averaging $\Reach$ and $\Path$}
\label{sssec:balance risk}
As described above both $\Reach$ and $\Path$ have weaknesses.
We demonstrate in Section~\ref{sec:eval} that $\fashion$
outputs better solutions when considering both metrics.  We call the
weighted sum of these two functions $\Hybrid$.  
%

$\Reach$ isolates nodes that do not need to communicate.  When
deciding which of two flows to route, $\Reach$ may not make the right
decision as it cannot quantitatively distinguish between the risk of
these flows.  However, in this case $\Path$'s measure can break ties
as it includes probability of exploitation while $\Reach$ does not.
 
$\Path$ is effective at isolating nodes that have a high probability
path to them.  Yet, this measure does not account for the
other nodes compromised ``on the way'' to the target node.  By
minimizing the total weighted reachable set using $\Reach$ this
weakness is partially mitigated.  Similarly, if two paths have similar
probability but one contains multiple AND prerequisites, it may have a
larger reachable component, enabling $\Reach$ to again break ties.

%


\section{Optimization Model}
\label{sec:model}
This section highlights the content and structure of the optimization
model used to obtain network configurations that uphold a balance
between functionality and security.  
The optimization model is a \emph{binary integer programming} (BIP)
model as all decision variables are binary. The model contains two
components dedicated,  respectively,  to the modeling of the data network
and its job as a carrier for data flows and to the representation of
the attack graph and the modeling of induced risk measures \Reach{}
and \Path{}. 
The core decision variables fall in two categories. 

First, Boolean variables model the routing decisions of the data flows
in the network. Such a variable is associated to each network link
and subjected to flow balance equations as well as
link capacity constraints. They also contribute to 
the functional rewards (in the objective) associated to the deliveries
of the flow values.  

Second, Boolean variables are associated to the deployment of
counter-measures in the network.  In this work, counter-measures
are routing a flow to a firewall (rather than its destination). 
Auxiliary Boolean variables facilitate the
expression of the model and setup \emph{channeling constraints} that
tie the attack graph model to the network routing model so that
routing as well as blocking decisions are conveyed to the attack graph
part of the model and  result in severing arcs that express pre-conditions of
exploits. 

While the constraints devoted to capturing reachability, i.e., \Reach{},
are relatively straightforward, the modeling of the most effective
path, i.e., \Path{} is more delicate for two reasons. First, it requires
the use of products of probabilities held in variables
yielding a non-linear formulation. Thankfully, that challenge can be
side-stepped by converting those products into sums with a logarithmic
transform. 
Second, it delivers a $\min-\max$ problem that needs to be dualized to
recover a conventional minimization.


\subsection{Full Model}
\label{sec:full model}
\newcommand{\B}{\{0,1\}}
\newcommand{\pred}{\delta^{-}}
\newcommand{\suc}{\delta^{+}}
\newcommand{\cost}{\textsf{cost}}
\newcommand{\val}{\textsf{val}}
\newcommand{\tsf}[1]{\textsf{#1}}
\noindent
In binary integer programming, 
the four primary components are Inputs, Variables, Constraints, and an Objective 
function. These are listed below.

\subsubsection{Inputs} 

$\mathcal{R}$ -- the set of SDN devices (routers/firewall).\\
$\mathcal{H}$ -- the set of hosts (machines) on the network. \\
$\mathcal{G} \subset \mathcal{R}$ -- the set of external gateway devices in the network.\\
$\mathcal{D} = \mathcal{R} \cup \mathcal{H}$ -- the set of all network devices. \\
$\mathcal{L} \subset D \times D$ -- the set of all network links. \\
$\mathcal{T}$ -- the set of traffic types. \\
$\mathcal{F}$ -- the set of tuples $(h,k,t) \in \mathcal{D} \times \mathcal{D} \times 
\mathcal{T}$
defining desired traffic flows of type $t$ from source device $h$ to
sink device $k$. \\
$s^*$ -- an artificial node in the network used as the global source of all flows.\\
$t^*$ -- an artificial node in the network used as the global destination of all 
flows.\\
$\src$ -- an artificial node in the attack graph used as the global starting point 
of the attacker. \\
$\Capab$ -- the set of capabilities an attacker could gain on devices in $\mathcal{D}$.\\
$\start$ -- the set of starting capabilities for an attacker. \\
$\ExploitN$ -- the set of exploits based on network connections. \\
$\ExploitV$ -- the set of exploits based on network vulnerabilities. \\
$\ExploitAND$ -- the set of \texttt{AND} exploits.\\
$\ExploitOR$ -- the set of \texttt{OR} exploits.\\
$\Exploit = \ExploitN \cup \ExploitV = \ExploitAND \cup \ExploitOR$ -- the set of all 
exploits. \\
$Q(l) : \mathcal{L} \rightarrow \mathbb{R}$ -- the capacity of link $l$. \\
$K(i) : \mathcal{R} \rightarrow \mathbb{R}$ -- the capacity of network device $i$. \\
$\tau(f) : \mathcal{F} \rightarrow \mathcal{T}$ -- yields the traffic type of flow $f$. \\
$\tau(\exploit) : \ExploitN \rightarrow \mathcal{T}$ -- yields the traffic type of 
network exploit $\exploit$.\\
$p(\exploit) : \Exploit \rightarrow [0,1]$ -- the probability of success for a given 
exploit.\\
$\pred(i) : \mathcal{D} \rightarrow 2^\mathcal{D}$ -- the set of vertices with outbound 
arcs leading to vertex $i$. \\
$\suc(i) :  \mathcal{D} \rightarrow 2^\mathcal{D}$ -- the set of vertices with inbound 
arcs originating at vertex $i$.\\
$\Pred(n) : N \rightarrow 2^N$ -- set of the predecessors of $n$ in the attack graph.\\
$\Succ(n) : N \rightarrow 2^N$ -- set of the successors of $n$ in the attack graph.\\
$\tsf{src}(f) \in \mathcal{D} $ -- the network device source of flow $f$.\\
$\tsf{dst}(f) \in \mathcal{D} $ -- the network device destination of flow $f$. \\
$\tsf{q}(f) : \mathcal{F} \rightarrow \mathbb{R} $ -- the quantity of data attributed to 
flow $f$.\\
$\val(f) : \mathcal{F} \rightarrow \mathbb{R}$ -- the value of flow $f$. \\
$\tsf{dev}(c) : \Capab \rightarrow \mathcal{D}$ -- yields the device in a capability 
node.\\
$\tsf{dev}(\exploit) : \ExploitN \rightarrow \mathcal{D}$ -- yields the destination 
device of a network-connection-based exploit.\\
$\cost(l) : \mathcal{L} \rightarrow \mathbb{R}$ -- the cost of an arc in the network. \\
$\Impact(c) : \Capab \rightarrow \mathbb{R}$ -- the impact of attack gaining capability 
$c$. \\

\subsubsection{Variables}
\noindent
$\rho_{i,j}^{f} \in \B$ -- for every $f \in \mathcal{F}$ and every $(i,j) \in 
\mathcal{L}$,
indicates whether link $(i,j)$ carries flow $f$. \\
$b_{i}^{f} \in \B$ -- for every $f \in \mathcal{F}$ and every $i \in \mathcal{R}$,
indicates whether flow $f$ is blocked by a firewall at network device $i$. \\
$w_{i}^{f} \in \B$ -- for every $f \in \mathcal{F}$ and every $i \in \mathcal{R}$,
indicates whether there is a firewall filtering flow $f$ at network device $i$. \\
$v_{i}^{t} \in \B$ -- for every $t \in \mathcal{T}$ and every $i \in \mathcal{R}$,
indicates whether there is a firewall blocking traffic type $t$ at network device $i$. \\
$W_{i} \in \B$ -- for every $i \in \mathcal{D}$, indicates whether there is any kind
of firewall present at network device $i$. \\
$N_{k,h}^{t} \in \B$ -- for every $(k,h) \in ( \mathcal{H} \cup \mathcal{G} )  \times
\mathcal{H}$  and every $t \in \mathcal{T}$, indicates whether there is an active network
connection between devices $h$ and $k$. \\
$\beta_{i}^{f} \in \B$ -- for every $i \in \mathcal{R}$ and every
$f \in \mathcal{F}$, indicates whether network device $i$ blocks flow
$f$ using a firewall specific to traffic
type. \\
$\theta_{i}^{f} \in \B$ -- for every $i \in \mathcal{R}$ and every $f \in \mathcal{F}$,
indicates whether network device $i$ receives flow $f$. \\
%
$r_{n} \in \B$ -- for every $n \in \Capab \cup \Exploit$, indicates whether node $c$ is 
reachable
from the root of the attack graph. \\
$a_{i,j} \in \B$ -- for every $(i,j) \in E$, indicates whether arc $(i,j)$ is present
(i.e., not cut) in the attack graph. \\
$x_{i,j} \in \B$ -- for every $(i,j) \in E$, indicates whether an attacker would be
able to traverse arc $(i,j)$, i.e., $r_{i} \land a_{i,j}$. \\
%

\subsubsection{Functionality Constraints }
 The first part of the model captures the networking side of
the model, including how to route flows, respect capacities of devices
and how to block specific flows. 
\begin{equation}
  \label{eq1}
  \rho^f_{s^*,src(f)} = 1, 
  \:\: \forall f \in \mathcal{F}
\end{equation}
Equation~\ref{eq1} forcibly spawns each flow in the network at its network machine 
source. 
\begin{equation}
  \label{eq3}
  \beta_{i}^{f} = \theta_{i}^{f} \land v_{i}^{\tau(f)},
  \:\: \forall f \in \mathcal{F}, \forall i \in \mathcal{R}
\end{equation}
\begin{equation}
  \label{eq4}
  b_{i}^{f} = w_{i}^{f} \lor \beta_{i}^{f},
  \:\: \forall f \in \mathcal{F}, \forall i \in \mathcal{R}
\end{equation}
Equation~\ref{eq3} models that a flow $f$ is blocked at device $i$
if it reached device $i$ and a firewall rule blocked traffic of the
type carried by $f$. Equation~\ref{eq4} is similar, but states that a
flow $f$ is blocked if there is a firewall rule specifically designed
to drop flow $f$ or if it was blocked by a generic firewall type
rule. 
\begin{equation}
  \label{eq5}
  \theta_{i}^{f} = \bigvee\limits_{j \in \pred(i)} \rho_{j,i}^{f},
  \:\: \forall f \in \mathcal{F}, \forall i \in \mathcal{R}
\end{equation}
Lastly, equation~\ref{eq5} states that a flow $f$ arrives at device $i$ if
one of the inbound arcs into $i$ carries flow $f$. 
\begin{equation}
  \label{eq6}
  \sum\limits_{j \in \pred(i)} \rho_{j,i}^{f} = \sum\limits_{k \in \suc(i)} 
  (\rho_{i,k}^{f} + 
  \beta_{i}^{f}),
  \:\: \forall f \in \mathcal{F}, \forall i \in \mathcal{R}
\end{equation}
\begin{equation}
  \label{eq7}
  \sum\limits_{j \in \pred(i)} \rho_{j,i}^{f} = \sum\limits_{k \in \suc(i)} 
  \rho_{i,k}^{f},
  \:\: \forall f \in \mathcal{F}, \forall i \in \mathcal{H}
\end{equation}
Equations~\ref{eq6} and~\ref{eq7} are the flow balance equations that
dictate that every inbound flow must also be outbound, unless it was
blocked. The equations are slightly different for internal SDN devices
(equation~\ref{eq6}) and hosts that are not permitted to route
traffic (equation~\ref{eq7}). 
\begin{equation}
  \label{eq8}
  \rho_{i,j}^{f} = 0,
  \:\: \forall f \in \mathcal{F}, \forall i \in \mathcal{H} \setminus 
  \{\tsf{src}(f),\tsf{dst}(f)\}, \forall j \in \suc(i)
\end{equation}
\begin{equation}
  \label{eq9}
  \rho_{\tsf{dst}(f),j}^{f} = 0,
  \:\: \forall f \in \mathcal{F}, \forall j \in \suc(i) \setminus \{t^*\}
\end{equation}
Equations~\ref{eq8} and~\ref{eq9} states that hosts can only forward traffic to the
flow sink by preventing the use of any other outbound arc. 
%
%
%
\begin{equation}
  \label{eq11}
  \sum\limits_{f \in \mathcal{F}} (\rho_{i,j}^{f} \cdot q(f)) \leq Q(i,j),
  \:\: \forall (i,j) \in \mathcal{L}
\end{equation}
%
%
%
\begin{equation}
  \label{eq12}
  \sum\limits_{f \in \mathcal{F}} \sum\limits_{k \in \pred(i)} (\rho_{k,i}^{f} \cdot 
  q(f)) +
  \sum\limits_{f \in \mathcal{F}} \sum\limits_{j \in \suc(i)} (\rho_{i,j}^{f} \cdot q(f)) 
  \leq K(i),
  \:\: \forall i \in \mathcal{R}
\end{equation}
Equation~\ref{eq11} simply models the bounds on link
capacities. Equation~\ref{eq12} plays a similar role for the device
capacities. 
%
%
\begin{equation}
  \label{eq15}
  W_{i} = \bigvee\limits_{f \in \mathcal{F}} w_{i}^{f} 
  \vee
  \bigvee\limits_{t \in \mathcal{T}} v_{i}^{t} \:\forall i \in \mathcal{R}
\end{equation}
Equation~\ref{eq15} models the existence of a blocking firewall of any
kind at SDN device $i$. 

\begin{equation}
  \label{eq22}
  N_{k,h}^{t} = 
  \begin{cases} 
    \rho_{h,t^*}^f & \mbox{ if } \langle k,h,t\rangle \in \mathcal{F} \\
      0 &\mbox{otherwise}
  \end{cases} ,
\end{equation}
\begin{equation*}
  \:\: \forall (h,k) \in ( \mathcal{H} \cup \mathcal{G} ) \times \mathcal{H},
  \forall t \in \mathcal{T}
\end{equation*}
Equation~\ref{eq22} models the existence of an active network connection between
devices $k$ and $h$.

Now, using the above constraints, we will derive a functionality score $\mathcal{O}_f$ for the network configuration. Looking ahead, this score will 
be used in defining the overall objective for the model.
\begin{align}
    \mathcal{O}_f &=\:\: \alpha_{0} \sum\limits_{f \in \mathcal{F}} -\rho_{dst(f),\sink} 
    					 	\cdot \tsf{val}(f) \nonumber \\
    					 &+ 
                        \alpha_{1} \sum_{f \in \mathcal{F}}
                		   \sum_{(i,j) \in \mathcal{L}} \rho_{i,j}^{f} \cdot 
                       		\tsf{cost}(i,j)\label{functObj}
\end{align}

The first term yields a credit for routing good flows through the network from their source to their
destination and is scaled by the assigned value of the flow. The second term describes the
cost incurred by routing flows though the network based on the cost of the links use to 
carry each flow. 

Lastly, we will use the functional constraints to derive a score for the cost of deploying network defenses, $\mathcal{O}_d$. 
This score will also appear in the objective function as 
part of the overall security cost.
\begin{align} \label{defObj}
\mathcal{O}_{d} = \alpha_2 \sum_{i} w_{i} + \alpha_3 \sum_{i} \sum_{t} v_{i}^{t} \:\: +
\alpha_4 \sum_{i} W_{i}
\end{align} 

The first two terms describe the cost paid for deploying flow-specific and traffic-specific
firewalls, respectively. The third term gives a penalty per unique network device
deploying any kind of firewall and its a way to reduce network complexity by encouraging
the concentration of multiple firewalls to a few devices. 

%
%

\subsubsection{Security Constraints}
Here we will lay out the necessary constraints to calculate the \Reach{} and \Path{} attack graph metrics, which will be needed 
in the formulation of the objective function of our model.

\paragraph{\Reach{} Measure} This part of the model is devoted to which 
capabilities an attacker can reach from the root of the attack graph. It creates channels between variables of the 
networking model and variables of the attack graph. It also models the semantics of 
{\tt  AND} and {\tt OR} nodes as well reachability within the attack graph.
\begin{equation}
  \label{eq16}
  r_{c} = \bigvee\limits_{i \in \Pred(c)} x_{i,c},
  \:\: \forall c \in \Capab
\end{equation}
\begin{equation}
  \label{eq17a}
  r_{\exploit} = \bigwedge\limits_{i \in \Pred(\exploit)} x_{i,\exploit},
  \:\: \forall \exploit \in \ExploitAND
\end{equation}
\begin{equation}
\label{eq17b}
r_{\exploit} = \bigvee\limits_{i \in \Pred(\exploit)} x_{i,\exploit},
\:\: \forall \exploit \in \ExploitOR
\end{equation}
Equation~\ref{eq16} models an {\tt OR} node. Namely, it states that
the capability $n$ is enable if the adversary can traverse at least
one inbound arc (coming from an exploit). Equation~\ref{eq17a} models
an {\tt AND} node. Namely, it states that the exploit $n$ is enabled
provided that \emph{all} inbound arcs can be traversed by the
adversary. Equation~\ref{eq17b} models an \texttt{OR} node in a similar fashion.
\begin{equation}
  \label{eq18}
  x_{i,j} = r_{i} \bigwedge a_{i,j},
  \:\: \forall (i,j) \in E
\end{equation}
Equation~\ref{eq18} enables an arc $(i,j)$ in the attack graph if its
source is reachable and the arc was not cut by $a_{i,j}$. 
\begin{equation}
  \label{eq19}
  a_{i,j} = 1,
  \:\: \forall (i,j) \in \ExploitV
\end{equation}
\begin{equation}
  \label{eq20}
  r_{\src} = 1
\end{equation}
Equations~\ref{eq19} state that arcs that are incident on a vulnerability-based exploit
$x$ are always enabled while equation~\ref{eq20} states that the attacker's starting
point $\src$, which has arcs leading to every $n \in \start$, is always reachable.
\begin{equation}
  \label{eq23}
  a_{c,\exploit} = N_{\tsf{dev}(c),\tsf{dev}(\exploit)}^{\tau(\exploit)},
  \:\: \forall c, \exploit \textrm { s.t. } (c,\exploit) \in E
\end{equation}
Equation~\ref{eq23} is a channeling constraint connecting the
presence of an network-connection-based arc in the attack graph to reachability in the
network model. 
Using the constraints just defined, we can formalize a security cost based on attack graph reachability, $\mathcal{O}_r$, 
that will appear in the objective function.
\begin{align} \label{reachObj}
\mathcal{O}_{r} = \sum_{n \in \Capab} r_n \cdot \Impact(n)
\end{align} 

Here we consider which capabilities an attacker can reach and use the associated impact of them achieving that 
capability as a penalty.

\paragraph{\Path{} Measure}
This part of the model focuses on capturing an attacker's \textit{most effective attack 
path}, which is formulated in Equation~\ref{pathRisk}, and embedding 
it within a minimization. To model this 
in a binary integer program, we use the strategy presented by Khouzani \emph{et al.} 
\cite{Khouzani2019a}. First, we note that Equation~\ref{pathRisk} is equivalent to 
the following.
\begin{align}
\label{maxProd}
& \max \limits_{\gamma} \prod \limits_{e \in E} 
	(\gamma_{e} p(t(e)) + 1 - \gamma_{e}) 
 \textrm{ s.t. } \gamma_{e} \in \{0,1\} \:\: \forall e \in E, \\
  &\sum \limits_{e:h(e) = i} \gamma_{e} - \sum \limits_{e:t(e) = i} \gamma_{e} =
	\begin{cases}\label{flowCons}
	1 \:\: & \mbox{if } i = \sink \\
	-1 \:\: & \mbox{if } i = \src \\
	0 \:\: & \forall i \in \Capab \cup \Exploit \\
	\end{cases}
\end{align}
Here $\gamma_{e}$ is a binary decision variable that indicates whether an arc $e$ is on 
the most 
effective attack path. Equation~\ref{flowCons} enforces that the arcs chosen will form a 
path from $\src$, the global attack graph starting point, to $\sink$, the global attack 
graph target. Here $h(e)$ and $t(e)$ 
denote the head $j$ and tail $i$ of an arc $(i,j)$. And since the 
objective function maximizes the probability of the taken path, as the product scales 
with an arc's probability if the arc is taken and 1 otherwise, this formulation is 
equivalent to equation~\ref{pathRisk}. However, as written, this formulation is not 
linear, as there is a product of variables in the objective function, and thus cannot be 
directly incorporated into a linear binary integer program. We can address this concern by 
composing the above objective with the monotonic function $\log(\cdot)$.
$$
\log \prod \limits_{e \in E} (\gamma_{e} p(t(e)) + (1-\gamma_{e})) = 
\sum \limits_{e \in E} \log(\gamma_{e} p(t(e)) + (1-\gamma_{e}))
$$
And considering that $\gamma_{e} \in \{0,1\}$, the above equation can be further 
simplified to $\sum \limits_{e \in E} \gamma_{e} \log(p(t(e))$. With this we now 
have the following optimization problem to model the most effective attack path.
\begin{align}
&\max\limits_{\gamma} \sum\limits_{e \in E} \gamma_{e} \log (p(t(e))) 
\textrm{ s.t. } \gamma_{e} \geq 0 \:\: \forall e \in E \\
\nonumber
\sum \limits_{e:h(e) = i} &\gamma_{e} - \sum \limits_{e:t(e) = i} \gamma_{e} =
\begin{cases}\label{flowCons2}
1 \:\: & \mbox{if } i = \sink \\
-1 \:\: & \mbox{if } i = \src \\
0 \:\: & \forall i \in \Capab \cup \Exploit \\
\end{cases}
%
\end{align}
%
%
Now we have obtained a linear formulation for the most effective attack path, but there 
is one more issue to address. The formulation is a maximization that we will be 
minimizing when synthesizing network configurations. Formulating this problem as a 
min-max is not desirable, so instead of solving the above maximization problem to find 
the most effective attack path, we will instead solve its \textit{dual}, which is a 
minimization problem with the same optimal solution. The dual model is given below.
\begin{align}
& \min \limits_{y} y_{\src} - y_{\sink} \\
\nonumber
\textrm{s.t.} \:\: & y_{t(e)} - y_{h(e)} \geq \log(p(t(e))) \:\: \forall e \in E
\end{align}
Note that because of the ability to deploy network defenses, we can apply $\log(\cdot)$ 
to equation~\ref{probRed} to obtain:
$$
\log(p(t(e))) = (1 - x) \log(p(t(e))) + x \log(\epsilon).
$$
Since we apply $\log$ to the probabilities, we cannot have an exploit with a probability 
of zero. Thus we use a very small number $\epsilon$ to model the scenario where network 
defenses reduce the likelihood of an exploit to zero. In particular, this is how 
completely severing host-to-host communications in the network can cut arcs incident to 
network reachability exploits in the attack graph.
Now that we have formulated a minimization problem here, it 
can be incorporated directly into the objective function of our overall minimization 
problem as the security cost due the most effective attack path, $\mathcal{O}_p$,
\begin{align} \label{pathObj}
\mathcal{O}_{p} = y_{\src} - y_{\sink}
\end{align} 

\subsubsection{Objective} 
The objective function in this model is comprised of two components, a functionality cost
and a security cost. The overall functionality cost is given directly by $\mathcal{O}_f$ as seen in 
Equation~\ref{functObj}.
We will define the overall security cost $\mathcal{O}_s$ as a linear combination of the security costs 
laid out in Equations~\ref{defObj},\ref{reachObj},\ref{pathObj}.
$$
\mathcal{O}_s = \beta_0 \mathcal{O}_d + \beta_1 \mathcal{O}_r + (1-\beta_1) \mathcal{O}_p .
$$
The cost of deploying network defenses is application dependent so we do not vary $\beta_0$ in our experiments, setting it to a constant of $\beta_0=1$.
One can vary $\beta_1$ to have the framework 
prefer one risk metric over the other. 
Finally, the overall objective that dictates a balance between functionality and security is a convex combination of 
the functionality and security costs:
\begin{equation}
  \label{objective}
  \min \:\: \alpha \mathcal{O}_f + (1-\alpha) \mathcal{O}_s
\end{equation}
Here we have the ability to vary $\alpha$ to influence the framework to produce 
network configurations that favor functionality over risk or vice versa.
Two key parameters of the optimization model are, therefore,
$\alpha$ and $\beta_1$ that control the functionality-risk tradeoff
and the balance between \Path{} and \Reach{} measures. 

\subsection{Extending to an Online Setting}
  \label{model-extension}
  As stated, $\fashion$ is suitable for deriving a network configuration that
  satisfies long-standing, or static, flow demands. However, in an online
  setting it is common for demands and exploits (discovery or patching) to
  change over time. In these cases, it is important to have an optimization
  framework that is robust in the wake of perturbations to input data.

  When updating a current network configuration, $\fashion$ should
  develop configurations that are functional, secure, \textit{and} similar in
  routing to the current setup. This would prevent a major routing overhaul
  being caused by the addition of a single nascent demand. This
  section extends $\fashion$ to take the current network state along with new
  demands/exploits as input and generate updated configurations.
\subsubsection{Extension Inputs}
$\tsf{sol}_{\rho} : \mathcal{F} \times \mathcal{L} \rightarrow \{0,1\}$ -- the value of $\rho^f_{i,j}$ in 
the current network configuration. \\
$\tsf{sol}_{b} : \mathcal{F} \times \mathcal{R} \rightarrow \{0,1\}$ -- the value of $b^f_{i}$ in 
the current network configuration. \\
$\tsf{sol}_{w} : \mathcal{F} \times \mathcal{R} \rightarrow \{0,1\}$ -- the value of $w^f_{i}$ in 
the current network configuration. \\
$\tsf{sol}_{v} : \mathcal{T} \times \mathcal{R} \rightarrow \{0,1\}$ -- the value of $v^t_{i}$ in 
the current network configuration. \\
\subsubsection{Extension Variables}
$\zeta$ -- the number of variable assignment differences between the previous
and current configurations.
\subsubsection{Extension Constraints}
The only constraint needed for the extension is one to calculate the number of
variable assignment differences between the previous and current solutions. It
is presented below in Equation~\ref{diff-calc}.
\begin{align}
  \label{diff-calc}
\nonumber
\zeta = \ \ & \sum\limits_{f \in \mathcal{F}} \sum\limits_{(i,j) \in \mathcal{A}} | \tsf{sol}_{\rho}(f,i,j) - \rho^f_{i,j} | \\
\nonumber
      &+ \sum\limits_{f \in \mathcal{F}} \sum\limits_{i \in \mathcal{R}} | \tsf{sol}_{b}(f,i) - b^f_{i} | \\ 
      &+ \sum\limits_{f \in \mathcal{F}} \sum\limits_{i \in \mathcal{R}} | \tsf{sol}_{w}(f,i) - w^f_{i} | \\ 
\nonumber
      &+ \sum\limits_{t \in \mathcal{T}} \sum\limits_{i \in \mathcal{R}} | \tsf{sol}_{v}(t,i) - v^t_{i} | 
\end{align}

\subsubsection{Extension Objective}
In order to guide the model towards configurations similar to the current one,
the number of variable assignment differences is directly appended to the usual
objective function. The complete objective function for the online FASHION
variant \rev{is} shown in Equation~\ref{ext-obj}.
\begin{equation}
  \label{ext-obj}
  \min \:\: \alpha \mathcal{O}_f + (1-\alpha) \mathcal{O}_s + \zeta
\end{equation}




\section{Evaluation}
\label{sec:eval}

In this section we show the efficacy and efficiency of $\fashion$. The goal is
to understand (i) whether $\fashion$ produces configurations that effectively
balance functionality and security and (ii) whether $\fashion$ produces
configurations quickly, what time scale of events can $\fashion$ respond to?
\subsection{Experimental Setup}
\label{ssec:exp_setup}

\begin{table*}[t]
\centering
\begin{tabular}{rrrrr ||r rrr|| r r r r}
  \hline
 \multicolumn{5}{c||}{Network Topology} & \multicolumn{4}{c||}{Network Traffic} & \multicolumn{4}{c}{Vulnerabilities}  \\ 
  \hline
 pod &  Devices & Hosts & Switch & Links &  \multicolumn{2}{c}{\#traffic} & \multicolumn{2}{c||}{\#flows}   &  \multicolumn{2}{c}{Exploits} &  \multicolumn{2}{c}{AG arcs}  \\ 
  &  & &  &  & min & max & min & max & min & max & min & max \\ \hline
2 & 10 & 4 & 6 & 9 & 1 & 3 & 8 & 80 & 1 & 10 & 18 & 165\\ 
4 & 37 & 16 & 21 & 52 & 1 & 3 & 32 & 320 & 1 & 40 & 258 & 2350 \\ 
6 & 100 & 54 & 46 & 171 & 1 & 3 & 108 & 1080 & 5 & 108 & 2928 & 26463 \\ 
8 & 209 & 128 & 81 & 400  & 1 & 3 & 256 & 2560 & 12 & 250 & 16422 &  147627\\ 
10 & 376 & 250 & 126 & 775  & 1 & 2 & 500 & 1500 & 12 & 400 & 62537 &  250765\\ 
12 & 613 & 432 & 181 & 1332 & 1 & 3 & 866 & 866 & 21 & 105 & 186682 &  1677306\\ 
   \hline\\
\end{tabular}	

\caption{Benchmark data, 649 instances.  The number of links represents the number of bi-directional links in the network. The $\#$traffic column represents the number of distinct traffic types.}
\label{tab:benchmarks}
\vspace{-.2in}
\end{table*}

To the best of our knowledge, no standard benchmarks for attack graphs exist.
For the purpose of this research, we created a benchmark suite with over 600
instances that models a data center topology and its traffic patterns and
utilization rates, along with a realistic representation of dispersed network
vulnerabilities. A high level breakdown of the benchmark characteristics can be
found in Table~\ref{tab:benchmarks}. The evaluation is made on a Linux machine
with an Intel Xeon CPU E5-2620 2.00 GHz and 64GB of RAM. The model
implementation was built in Python 3.7 using the Gurobi optimization
library~\cite{gurobi}.

\paragraph{Network Topology} The generated instances are of the popular
Clos~\cite{Clos1953} style network topology
Fat-tree~\cite{Al-Fares:2008:SCD:1402958.1402967} and are representative of a
cloud data center. Fat-tree features an expanding pod structure of
interconnected and tiered switches providing excellent path redundancy. The
topology is designed to deliver high bandwidth to many end devices at moderate
cost while scaling to thousands of hosts. Switch and link capacities in all
benchmarks are 1GBps. The benchmarks include small to medium sized instances.
The largest instance tested, includes 613 devices (hosts and SDN devices) and
1332 links between these.

\paragraph{Network Traffic} Network demand is modeled after
the recent Global Cloud Index (GCI) report~\cite{CISCO:2018} which provides
a global aggregated view of data centers.  The benchmarks include two
distinct traffic patterns: Internal at $70\%$ and External at $30\%$ (by combining  GCIs Inter-data center $15\%$ and Internet $15\%$). Research shows that there exists
heavy-tailed distributions for the volume and size of data
flows~\cite{Alizadeh2010}. There are generally small (1Mb-10Mb) and
large (100Mb-1Gb) sized flows with 90\% of the traffic volume being
small and 10\% large~\cite{Zhou2019}, all benchmarks follow this distribution.

Each flow is randomly labelled with a traffic type with number of types ranging
from 1 to 3, and number of flows per host varying in $\{1,3,5,10\}$. Each flow
is randomly labeled with a traffic type to account for the range of traffic such
as \texttt{HTTP,HTTPS,SMB}~\cite{Benson:2010:NTC:1879141.1879175}. For
considered instances, the number of traffic type varies from 1 to 3.

  One-third of the instances have 1 traffic type, one-third of the instances
  have 2 traffic types and the one-third of instances have three traffic types.
Each flow is assigned a flow value at random from the set $\{1,2,3,5,25\}$.
  Network utilization is impacted by several factors such as time of
  day~\cite{Delimitrou2012}, or application distribution~\cite{Kandula}
The number of flows per host is varied across benchmarks with steps
$\{1,3,5,10\}$ to vary network utilization~\cite{Delimitrou2012,Kandula}.

  As an illustrative example of the benchmark generation process consider an
  instance with 54 hosts and 10 flows per host resulting in $540$ bidirectional
  flows. Each bidirectional flow is first randomly assigned as internal ($70\%$)
  or external ($30\%$). Next, the source and destination are randomly selected,
  two distinct hosts for internal or one host and the Gateway switch for
  external. (Note, the Gateway is the demarcation point between the network and
  the Internet) Each flow is then assigned a size, traffic type and value based
  on the distributions provided above. Finally, each flow is duplicated
  reversing the source and destination to represent two-way traffic, resulting
  in $1080$ total flows.

 \paragraph{Vulnerabilities} Synthetic vulnerabilities are
injected on hosts within the network. The generation adopts several
components from the vulnerability model presented in the recent CVSS 3.1.
Base metrics focusing on exploitability and
impact~\cite{website:CVSS}. The percentage of \textit{exploitable
  hosts} ranges in \sloppy {$\{10\%,20\%,30\%,40\%,50\%\}$} and the
average number of vulnerabilities per host (1-5) drives the total
number of vulnerabilities injected. The number of 
vulnerabilities per host is representative of Zhang \emph{et al.}'s
findings~\cite{Zhang2014} from scans of publicly available VMs after
patching was performed.

  Each vulnerability has one or more prerequisite conditions and a single
  post condition of privilege escalation (if successfully exploited). Each
  vulnerability is uniformly assigned a score $[0,1]$ representing the
  probability of exploitation. Three privilege levels $\{0,1,2\}$ are assumed
  where 0 denotes networking reachability.

   Procedurally, the single precondition exploits are generated at random from
  selected exploitable hosts, each allowing the escalation of a single step in
  privilege level. The precondition of these single-precondition exploits have a probability of $.50$. If the required number of single precondition exploits
  exceeds the number of exploitable hosts then additional random hosts have
  vulnerabilities manually placed on them.

  Multi-precondition exploits are generated by selecting one prerequisite
  randomly from the pool of existing (single precondition) exploits and
  generating a new exploit which increases the privilege of one of the input
  exploits, the other, secondary prerequisites are selected randomly. The preconditions of these multi-precondition exploits have a probability of $.25$ each. 
  Importantly, the secondary prerequisite selection is restricted to currently
  \emph{reachable} exploits to ensure the attack graph has large connected
  components. The primary prerequisite (the privilege that will be escalated)
  selection is not restricted to currently achievable nodes.

  The impact of a successful exploit is reflective of the value of the
  threatened host. We uniformly assigned each host an integer in $[1,100]$ to
  represent its value to an organization. The impact of compromise of reaching a
  privilege level on a host is a percentage of this asset value. The quantities
  $[20\%,40\%,100\%]$ are used as the scaling factors for the three privilege
  levels $\{0,1,2\}$.

\subsection{Results}
\label{ssec:results}

We first focus on the basic model designed for an initial configuration, we then
consider the online setting in Section~\ref{sssec:online}.

\subsubsection{Evaluating $\fashion$'s Initial Configurations}
The discussion focuses on answering the three questions: 1) does $\fashion$
produce good configurations? 2) does it do so in a timely manner? and 3) can
$\fashion$ work in an online manner?
Answering the first question is slightly delicate because our security
optimization is using $\Reach$ and $\Path$ instead of using $\Risk$. Throughout
this section we only report on the security quality of the final configuration
with respect to $\Risk$ (we view this as our baseline metric). This metric
  was introduced by Frigault et al.~\cite{frigault2017measuring} for measuring
  risk in Bayesian attack graphs. The algorithm for computing $\Risk$ is too
slow to use in the optimization model but allows an effective check on the
quality of the solution a posteriori. In all results the functionality score is
the normalized value of the delivered traffic: it considers the total value of
traffic delivered when $\alpha~=~1$ (corresponding to the optimization
considering only functionality) as a functionality score of $1$ with the other
functionality scores normalized accordingly. $\Risk$ is normalized in a similar
way and computes the $\Risk$ value when $\alpha~=~1$ (no protections deployed)
and uses this as the denominator for other configurations. When $\alpha~=~1$
this corresponds to a baseline $\Risk$ for all trade-offs of the security model.
This is because the security model is inactive in the optimization.

\ifnum\lncs=0
\subsubsection{Does $\fashion$ produce good configurations?}
\label{sssec:heuristics work}
\fi
Section~\ref{sssec:balance risk} argues that combining the two
security models would produce better configurations than having
$\beta_1~=~0$ or $\beta_1~=~1$.  This was confirmed in our experiments.
%
%
Setting $\beta_1=1$ produced a meaningful trade-off between functionality
and security.  However, when we consider $0 < \beta_1 < 1$ in the
$\Hybrid$ manner, solution quality improves.  In all analyzed
solutions, varying $\alpha$ with just the $\Path$ measure active
($\beta_1 =0$) produced solutions that varied $\Path$ but not the actual
$\Risk$ (other than blocking the gateway).

Thus, the setting of $\beta_1 \in (0,1)$ seems crucial, but the
particular value within $(0,1)$ does not seem to have a substantial
effect on solution quality.  This would be the case if one model is
primarily being used as a tie-breaker for the other model.  However,
we cannot rule out that different settings of $\beta_1$ would be
preferable on different classes of networks and attack graphs.
For the remainder of the analysis, consider $\beta_1=.5$, i.e., $\Path$
and $\Reach$ are equally weighted.  

\paragraph{Monotonicity}
One can ask if $\fashion$ produces
solutions that trade off functionality and security.  As a reminder,
we use whitelist and source specific routing so setting $\alpha=1$
corresponds to the minimum risk that is achievable while routing all
desirable flows (assuming routing all flows is feasible within
bandwidth constraints).  Any further minimization of risk necessitates
decreasing functionality.  Furthermore, since we consider an external
attacker, blocking all flows at the entry point is always the solution
chosen when functionality is not considered (optimizing only over
risk).  Thus, every instance has functionality and risk of $0$ when
$\alpha~=~0$.  We remove this point from all analysis and consider
$\alpha~>~0$.  Furthermore, we normalize both risk and functionality
by $\alpha=1$ so we remove this point as well.

\begin{figure*}[t]
\centering
	\subfigure{
		\centering
		\includegraphics[width=.44\textwidth]{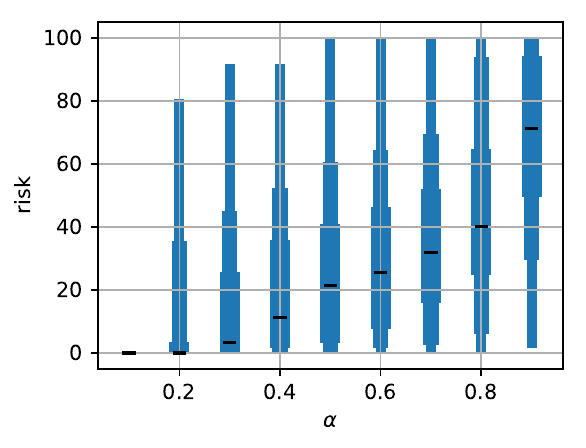}	
	}\hfill
	\subfigure{\centering
		\includegraphics[width=.44\textwidth]{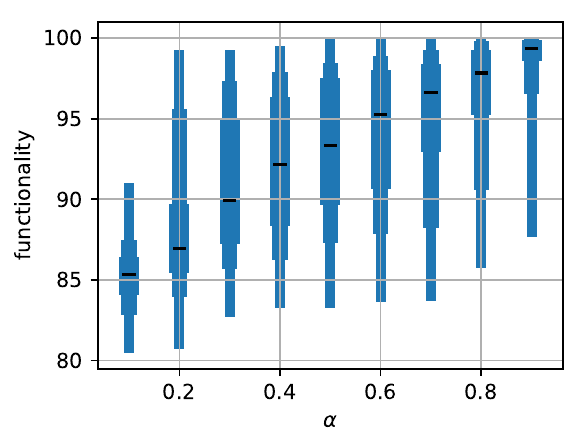}	
	}
	\vspace{-.1in}
\caption{Probability mass function of instances for normalized risk (left) and
normalized functionality (right).  Start of bar represents min, widening at 10\%
and 25\% with a line at 50\%, reducing width at 75\% and 90\%, stopping at the max
for each $\alpha$. In all cases, risk is $0$ when $\alpha=.1$.  Both scores are normalized by the setting when $\alpha = 1$. This corresponds to the security layer being inactive.  Functionality is measured directly using \fashion{}'s functionality objective, $\mathcal{O}_f$. Because our security layer approximates risk in an attack graph, here we directly report on Frigault et al.'s~\cite{frigault2017measuring} metric.  Our security measure $\mathcal{O}_s$ behaved monotonically as well.}

	\label{fig:EB}
\end{figure*}

Since $\fashion$ is approximating $\Risk$ it is possible for a decrease in $\alpha$ to lead to worse
risk and functionality. However, in all of our experiments, risk and
functionality were both monotonic for steps of $\alpha$ of size
$.1$.\footnote{We did observe small perturbations that violated
monotonicity if one considered steps of $\alpha$ of $.01$.}
We consider 212 benchmarks of pod size of 6, each having 54
hosts. Varying $\alpha$ for each benchmark we plot both the
functionality scores against $\alpha$ and also the risk scores against
$\alpha$. The functionality and risk scores are normalized as
previously stated.  Figure~\ref{fig:EB} demonstrates the probability
mass functions over every instance solution varying $\alpha=.1$ to
$\alpha=.9$ in steps of $.1$.
Functionality scores across all benchmarks are
relatively stable for $\alpha\ge .1$ while the risk scores vary
more.  (The two graphs have different $y$-axes.)

Importantly, at every point on this curve $\fashion$
is computing and outputting a corresponding network configuration.
Note that in addition to considering what flows to include, the
solution also describes how to route flows in a way that respects switch and link capacity.  Based on visual inspection we
classified our instances into three types of attack graphs.

\noindent
\textbf{Instances where exploitable hosts are critical}  In
such instances nodes with exploits serve as destinations in many of the
desired flows. In one generated pod 4 attack 
graph, a node with an exploit was the end point for a flow from $11$
of the $15$ other hosts.  ``Disconnecting'' this node from the network
required sacrificing many flows.  This yields a sharp functionality
vs. security trade-off. Note that such a graph can occur in practice when
many clients need to access a critical, vulnerable resource such as a database.   

\noindent
\textbf{Instances where exploitable hosts are isolated} In such instances flows with high value are mostly
distinct from exploitable nodes.  In one instance with $98$ total
flows, it is possible to achieve risk $0$ by only blocking $15$ flows.
Such instances can be seen as easy: the risky nodes are not crucial to
functionality. 

\noindent
\textbf{Instances with many trade-offs} In such
instances exploitable hosts  have a meaningful but not overwhelming value of flow. 
In these cases, \fashion{} can mitigate risk in two ways: by 
severing external connections to prevent an attacker from 
entering the network, or by severing internal connections 
to prevent an attacker from moving laterally through network.

To illustrate the changes that occur as one changes $\alpha$, we consider one pod 4 attack graph with four exploitable nodes.  
These four nodes are all 
involved in both external and internal flows. Here the 
external flows were typically of higher value than that of internal 
flows, making $\fashion{}$ sacrifice the internal flows for the sake 
of security at larger values of $\alpha$. However as $\alpha$ 
decreases, external flows begin to be blocked which allows for previously 
severed internal flows to be serviced once again, as they are no longer 
needed to prevent lateral movement since the attacker cannot necessarily 
enter the network through external gateways. Table~\ref{tab:tradeoffs} 
shows the balance of external and internal flows blocked in this 
instance. In this instance, the larger sets of blocked internal flows 
at smaller values of $\alpha$ were not supersets of smaller sets of blocked 
internal flows seen at larger $\alpha$ values.  This demonstrates an important
capability of $\fashion$: the ability to recognize defenses whose current marginal cost (to functionality)
exceeds their value (to security).
 
\begin{table}[tb]
\scriptsize
	\centering
	\begin{tabular}{|l|r r r r r r r r r|}
		\hline
		$\alpha$               & .9 & .8 & .7 & .6 & .5 & .4 & .3 & .2 & .1 \\
		\hline
		External Flows Blocked &  0 & 1  & 3  & 5  & 5  & 7  & 7  & 48 & 48 \\
		Internal Flows Blocked &  3 & 11 & 11 & 15 & 15 & 23 & 23 & 0  & 0  \\   
		\hline
	\end{tabular}
	\vspace{.05in}
	\caption{Number of external and internal flows blocked over varying values of 
		$\alpha$ on example instance.}
	\label{tab:tradeoffs}
\end{table}

\ifnum\lncs=0
\subsubsection{Does $\fashion$ produce configurations in a timely manner?}
\fi

\paragraph{Agility}To verify the scalability of \fashion{} and its ability to react to
short term events, it is valuable to assess performance as a function
of various input size parameters.
Experiments were done on Fat-tree networks with pod sizes 6, 8, 10, and 
12. 
Table~\ref{tab:podsRT} shows the model solve times when scaling the
number of flows per host.  In this
experiment, the sizes of 
pods, number of flows as well as the number of exploits per host were
increased. Rows labelled $25\%$, $50\%$ and $75\%$ report the
percentile breakdown of the runtime for the instances considered. The
columns vary the number of pods (6 to 12) as well as the number of
flows per hosts.  

The key observation here is that both the number of hosts and flows
contribute significantly to the solve time. This is not surprising as
the amount of hosts and flows affect the sizes of both the network and
the attack graph, resulting in a large increase to the model size.  We
note that the average runtime does not always strictly increase in
Table~\ref{tab:podsRT}. This is likely due to a smaller number of pod
10 \& 12 instances being tested due to solve time, contributing to
higher variance among their reported times.

\begin{table}[t]
        \tiny
	\centering
	\begin{tabular}{|l|rrr|rrr|rr|r|}
		\hline
	pod size              & \multicolumn{3}{r|}{6}     & \multicolumn{3}{r|}{8}  & \multicolumn{2}{r|}{10} & 12\\     
	
	end hosts             & \multicolumn{3}{r|}{54}     & \multicolumn{3}{r|}{128}  & \multicolumn{2}{r|}{250} & 432\\   
		SDN devices             & \multicolumn{3}{r|}{46}     & \multicolumn{3}{r|}{81}  & \multicolumn{2}{r|}{126} & 181\\     \hline
	\#flows per host & 3   & 5      & 10 & 3   & 5 & 10 &  1  & 3& 1\\
	\hline
	min              & 13 & 25  & 65  & 152 & 466 & 1069 & 1650 & 1540 & 1256\\
	25\%             & 16 & 38  & 120 & 195 & 700 & 1545 & 2399 & 1608 & 1469\\
	50\%             & 19 & 43  & 151 & 250 &  763 &1921 & 2686 &  1733 & 1853\\
	75\%             & 23 & 53  & 201 & 309  & 896 & 2259  & 2845 & 1990 & 2212\\
	max              & 87 & 289 & 2316 & 607 & 2787 & 2825  & 2979 &2308 & 2977\\
  average          & 22 & 57  & 205 & 263 & 863  & 1937  & 2576 & 1836 & 1940\\
	\hline
	\end{tabular}
        \vspace{1mm}
	\caption{Solve Time in Seconds for Pod 6, 8, 10 and 12 Networks with Different Flows Per Host.  Objective is linear combination of functionality and security objectives.  Results with functionality weight at $.7$ and security weight at $.3$.}
        \label{tab:podsRT}
        \vspace{-.3in}
\end{table}

Another trend is that the number of exploits alone can
cause a rise in solve time, as it directly affects the size of the
attack graph. Yet, the volume of exploits does not have as dramatic an
impact as the number of hosts/flows. 

These results show that for networks of up to 128 hosts affected
by a substantial number of flows and exploits, the framework produces
optimal configurations within 3-7 minutes. 

Since optimization models are often applied to NP-hard problems, they eventually hit a \emph{knee} in the curve where they stop being tractable.  Based on our experiments in Table~\ref{tab:podsRT}, pod 12 solutions are approaching that \emph{knee}.  One may be able to improve scalability with a different optimization model.
As an alternative approach, Ingols et al. are able to scale to attack graphs with 40,000 hosts in a 
similar time period by introducing equivalence classes among hosts to reduce the 
size of the attack graph and achieve scalability~\cite{ingols2009modeling}. 
This technique can be applied in our setting as well.  

Homer et al. \cite{Homer2013} build probabilistic attack graphs on
networks with $100$ hosts, their attack graph generation takes between
1-46 minutes depending on the complexity of the exploit chains. Both
prior works generate attack graphs for \emph{static network configuration}
with no consideration of the functionality problem at all.

$\fashion$'s response time (on networks of this scale) allows
automated response to short term events such as publication of a
zero-day before a patch is available.  In an actual deployment, where
the model must be solved repeatedly over time as inputs slightly
evolve, the runtime can be drastically reduced when resolving the
model by priming the optimization with the solution of the previous
generation~\cite{Ralphs2006,pour2018}.

\noindent
\subsubsection{Evaluating $\fashion$'s Online Extension} 
\label{sssec:online}
In this subsection, we test the effectiveness of \fashion{}'s online extension
discussed in Section~\ref{model-extension}. We split our discussion into updates
of new required traffic and unknown vulnerabilities.

\paragraph{Additional Demand}
To test the effectiveness of the online extension, a series of instance
pairs were created for Fat-tree pod size 4,6, and 8 networks. Each pair
consisted of a normally generated, or \textit{orginal}, instance, and another
instance which augmented the original with a new 
flow demand.

$\fashion$ was first run on the original instances to generate the current
network configuration, $\mathsf{ConfigA}$. After this $\fashion$ was run twice
on the augmented instances, once without the online extension and once with the
extension. This process generated network configurations $\mathsf{ConfigB}$ and
$\mathsf{ConfigC}$, respectively. Additionally, when generating
$\mathsf{ConfigC}$ the solver was warm started with the solution found in
$\mathsf{ConfigA}$.  $\mathsf{ConfigB}$ is \fashion{}'s  solution for the
overall demand if consistency with the prior solution is not considered.  

In all of our tests (100 randomly generated configuration pairs per pod size),
$\mathsf{ConfigB}$ and $\mathsf{ConfigC}$ yielded the same objective score
(though usually with very different variable assignments due to Fat-tree's
redundancy), but $\mathsf{ConfigC}$ had no differences from $\mathsf{ConfigA}$
besides the nascent variables related to the new flows. That is, optimizing for
similarity to $\mathsf{ConfigA}$ caused no degradation to the objective value
and \fashion{} was able to produce updates for an SDN controller that only
specified how to deal with the change.

Therefore,  the online extension was able to generate updated
network configurations that both minimally differed from the current network
state \textit{and} achieved the same functionally and security that would be
found in running the framework from scratch. Lastly, when using a warm start,
$\fashion$'s solve time was reduced by a factor of two.

\paragraph{Additional Exploits} When a new exploit is found and added to $\fashion$'s 
input data, the severity of the exploit and its exploitability within the current 
network configuration determines how much change is required to sufficiently reduce 
risk. In many tests, the exploit generated and added to the network input was not exploitable 
in terms of the current configuration, requiring no change from $\fashion$. 

However, when the new exploit changes the attack graph, the resulting
configuration should change. To exercise this capability, we generate exploits that
are likely to be harmful in the current configuration. In these cases,
$\fashion$ mitigated risk by strategically placing firewalls while minimizing
the impact to any benign flows through small-scale rerouting. Typically the
altered flows have a common prefix and suffix compared to the original routes,
so the alteration would require updates to few SDN devices.

To illustrate, consider one pod 4 configuration. A severe privilege
escalation exploit was added (with respect to the initial configuration).
$\fashion$ was able to cutoff all external reachability to the affected device
by placing a single firewall. Interestingly, the firewall was placed in a core
router rather than in an edge router adjacent to the exploitable device. To deal
with this new firewall, 10 out of 95 benign flows were rerouted.

There were two additional internal flows destined for the vulnerable device
(which the attacker could have used to move laterally). The security layer made
the decision to firewall these flows to limit the attacker's ability to pivot.
One of the blocked flows took a very long route through the network before
eventually reaching the firewall. This odd-looking, circuitous route was chosen
to maintain a common routing prefix/suffix and result in fewer overall rule
changes pushed to SDN devices.

While these choices may be unintuitive for a network engineer, $\fashion$ was
able to analyze all of impacts of the firewall placement and made the best
decision in order to both reduce risk and prevent massive routing upheaval. One
could remove the similarity portion of the objective which would cause more
traditional, short routes to firewalls to be chosen for the terminated flows (at
the expense of more changes to SDN devices!).

\paragraph{Online Extension Summary} These results show that, within Fat-tree's highly redundant
structure, $\fashion$ is able to update current network configurations
to accommodate new demands/exploits with very little overhead. The overhead
comes from the model solve time (which is significantly reduced due to warm
starting) and editing old/pushing new rules to the SDN devices.
In many cases optimal solutions when considering similarity had the same
solution quality as those that did not consider similarity.
This means there would not be a need to edit existing rules within
the routing tables of network devices. One would only need to push a few new
rules to the relevant devices. At the same time, \fashion{}'s online extension
does effectively respond when the change impacts the overall risk to the
network.



\section{Conclusion}
\label{sec:conclusion}

There are three shortcomings to making
  attack graphs an actionable tool in a network administrator's toolbox. These
  are 1) fast evaluation, 2) balancing risk with functionality, and 3) quick
  deployment of changes. SDNs provide the promise for quick integration. The
  goal of \fashion{} is to address quick evaluation while considering
  functionality.

\ifnum\lncs=0
Configuring Software Defined Networks to maximize the volume of
customers data flows to and from servers while respecting device and
link capacities is a classic flow optimization problem. Protecting
such a network from adversaries attempting to exploit vulnerabilities
that plague specific devices and hosts is equally important to address
within organizations. Attack graphs are effective in modeling risk and
finding mitigations (defensive measures). 
Unfortunately, risk and functionality are antagonistic objectives and
optimizing one without caring for the other is unhelpful as it will
deliver extreme solutions that are impractical. 
This paper considers both challenges in a holistic fashion and
automatically computes new SDN configurations for network devices in
response to emergent changes in demand, component risk or exploit discoveries. The $\fashion$ framework models
the customer demands, network devices and link capacities. It also
captures two notions of risk, $\Path$ and $\Reach$ under an
attack-dependency graph model within the overall
optimization. The output from $\fashion$ includes routing decisions for
SDN devices as well as firewall mitigation decisions. 
\fi

\paragraph{Evaluating attack graphs quickly}
To underscore the need for faster evaluation of attack graphs, we benchmark the
prior Bayesian attack graph benchmark of Frigault et
al.~\cite{frigault2017measuring}. As a reminder, Frigault et al.'s metric is the
\emph{baseline} when assessing the security quality of solutions in
Figure~\ref{fig:EB}.

\begin{figure}[tb]
\begin{center}
\includegraphics[scale=0.45]{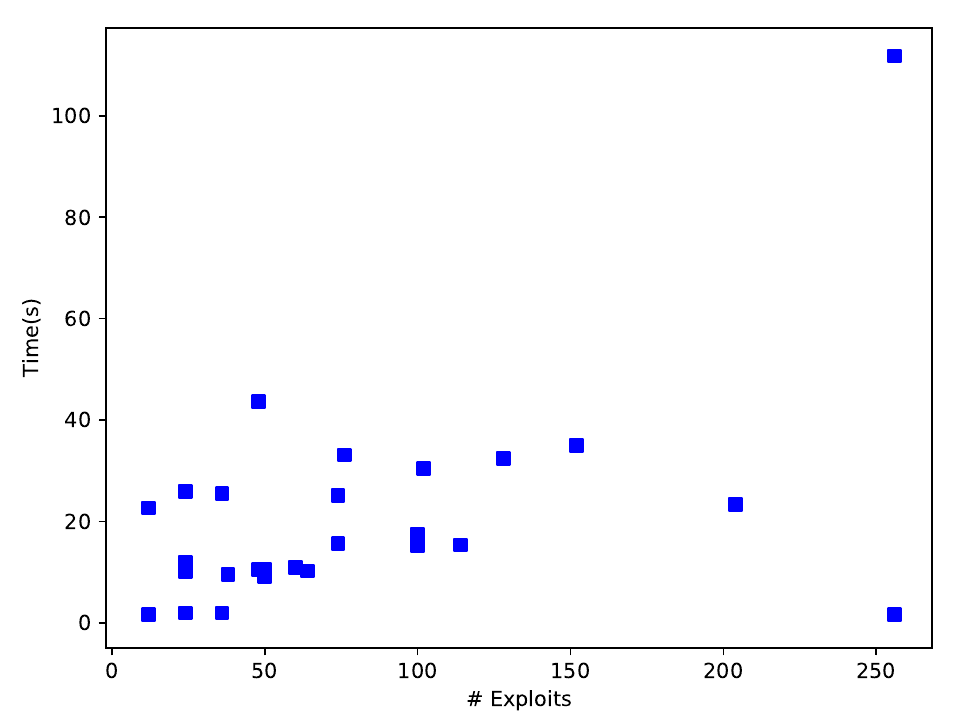}
\end{center}
\caption{$\Risk$ calculation
  time for output of $\fashion$ on pod $8$ instances using Python
  implementation of Frigault et al.'s algorithm~\cite{frigault2017measuring}.} 
\label{fig:lin}
\vspace{-.15in}
\end{figure}

We now ask if the algorithm of Frigault et al.~\cite{frigault2017measuring}
could be directly used in place of $\Reach$ and $\Path$. We created a Python
implementation of this algorithm~\cite{fashion-repo}, analyzing instances of 128
hosts, with 12 to 250 vulnerabilities. Figure~\ref{fig:lin} shows the time
required to compute the actual $\Risk$, with $\alpha=.7$ using our Python
implementation of the algorithm by Frigault et al.~\cite{frigault2017measuring}
of \textit{one} network configuration output by \fashion{}. Over this set of benchmarks,
the computation time for the risk values \emph{of a single fixed configuration}
can reach 100 seconds for 250 exploits. $\fashion$ inspects thousands of
configurations during its search. The time to evaluate the risk seems correlated
to the number of exploits, and the evaluation algorithms start to struggle even
with a relatively small number (e.g., 50) exploits.

The search space associated with this kind of problem is generally huge.
Directly specifying $\Risk$ would yield a highly non-linear, and thus likely
intractable, formulation. Enumerating configurations and computing the risk a
posteriori would be equally intractable. \emph{Linearization is therefore
  essential} in a responsive and practical framework.

  \paragraph{Incorporating Functionality}The paper demonstrates that $\fashion$
  can explore the trade-off between functionality and risk.

As stated, $\fashion$ optimizes over both objectives but the model can easily be
converted into one where either security or functionality is a \emph{constraint}
and the other objective is optimized. The average of $\Path$ and $\Reach$ is an
effective linearizable stand in for a risk calculation that is prohibitively
expensive to compute on the scale needed for a configuration search problem.
Interestingly, the novel hybrid risk model enables $\fashion$ to
overcome their respective weaknesses and produce better solutions.
Practically, $\fashion$ runs in matter of minutes on
networks of reasonable size (613 devices) and demonstrates potential for
scalability. Finally, the empirical results indicate that the
approximation adopted by $\fashion$ does not jeopardize key properties
such as monotonicity of functionality vs. risk. 

The $\fashion$ framework is a first step towards handling both
functionality and risk for short-term response while producing
consistent results with natural interpretations. 
Future directions include improving handling scalability of the
model size that currently depends on the number of network
links as well as the number of data flows and supporting a more varied set
of controls beyond routing and blocking. Additional future work is understanding if there are network settings that require a different weighting of the two measures of $\Path$ and $\Reach$. 
Lastly, evaluation of $\fashion$ configurations should continue using real data and real SDNs to understand emergent artifacts.

  \paragraph{Other settings} \fashion{}'s primary application is quick response
  to new security information in an SDN. However, attack graph analyses are
  notoriously complex. \fashion{}'s security layer is useful for network
  planning as well. \fashion{} allows an organization to consider ``what-if''
  queries for proposed changes in the network. One could consider the deployment
  of a new webserver on either {\tt Apache} or {\tt Nginx} (along with introduction of
  corresponding vulnerabilities) and understand how this would affect the
  overall risk posture of the network. Importantly, these analyses would
  consider this risk while maintaining network functional requirements. Once a
  small set of possible solutions has been identified, one can then validate
  \fashion{}'s output as described in the Introduction.

\section*{Acknowledgment}
\label{sec:ack}
\noindent
The authors thank the anonymous reviewers for their helpful insights. The
authors would also like to thank Pascal Van Hentenryck and Bing Wang for their
helpful feedback and discussions. The work of T.C., B.F., H.Z. and L.M. was
supported by the Office of Naval Research, Comcast and Synchrony Financial. The
work of D.C. is supported by the U.S. Army. The opinions in this paper are those
of the authors and do not necessarily reflect the opinions of the supporting
organizations.

\ifnum\lncs=1
\section*{Authors}
\emph{Devon Callahan} (devon.m.callahan.mil@mail.mil) is an Assistant Professor of Computer Science at the United States Military Academy, West Point. He is an Active Duty Army officer with over \rev{24} years of service including multiple combat tours. He received a B.S. degree from Methodist University in 2004, a M.S. from the University of South Carolina in 2009 and a Ph.D. from the University of Connecticut 2020. His research focuses on networks (computer and social) and systems security.

\emph{Timothy Curry} (timothy.curry@uconn.edu) is a PhD student at the University of Connecticut specializing in combinatorial optimization applied to security problems.  Tim received B.S.'s in mathematics and biological sciences from the University of Connecticut in 2014.

\emph{Hazel Davidson} (daniel.h.davidson@uconn.edu) is a software engineer at Microsoft.  Prior to joining Microsoft, she received her undergraduate degree from  the University of Connecticut.  Her research interests include programming languages and security.

\emph{Heytem Zitoun} (Heytem.zitoun@gmail.com) is a consulting researcher for Huawei Research Center located in Paris France. Prior to work as a Research Consultant, he was a Research Assistant at the University of Connecticut under Laurent Michel supervision. 
Heytem received a  master's degree in computer science at Universit\'{e} of Nice-Sophia Antipolis in 2014, and received a Ph.D. degree in 2018 from Universit\'{e}  C\^{o}te d'Azur. 
His research focuses mostly on Constraint Programming, Verification, and more generally AI techniques. 

\emph{Benjamin Fuller} (benjamin.fuller@uconn.edu) is an Associate Professor of Computer Science at University of Connecticut.  Prior to joining the University of Connecticut, he was a technical staff member at Massachusetts
Institute of Technology (MIT) Lincoln Laboratory from 2007 to 2016. 
 He received a B.S. degree
from Rensselaer Polytechnic Institute in 2006 and M.A. and
Ph.D. degrees from Boston University in 2011 and 2015, respectively. His
research focuses on applied cryptography.

\emph{Laurent Michel} (laurent.michel@uconn.edu) received the M.S. and Ph.D. degrees in computer science
from Brown University, Providence, RI, USA in 1996 and 1999, respectively.
He is the Synchrony Chaired Professor in Cybersecurity in the computer science and engineering department at the University of Connecticut, Storrs, CT, USA. His interests span combinatorial optimization with a particular emphasis on constraint programming, security and voting technology.
\fi




\bibliographystyle{ieeetran}
\bibliography{biblio}

\end{document}